\documentclass[11pt]{article}

\usepackage[a4paper,hmargin=1.0in,vmargin=1.0in]{geometry}
\usepackage[linktocpage=true,pagebackref=true,colorlinks,allcolors=blue,bookmarks=true]{hyperref}
\usepackage[numbers]{natbib}
\usepackage{amsthm,amsmath,amssymb,amsfonts,mathtools,sectsty,xcolor,setspace,xspace,bbm}

\newtheorem{theorem}{Theorem}
\newtheorem{lemma}{Lemma}[section]

\theoremstyle{definition}

\let\oldproofname=\proofname
\renewcommand{\proofname}{\rm\bf{\oldproofname}}

\numberwithin{equation}{section} 

\newcommand{\MyAbove}[2]{\genfrac{}{}{0pt}{}{#1}{#2}}
\newcommand{\eps}{\epsilon}
\newcommand{\ex}[1]{\mathbb{E}\left[#1\right]}
\newcommand{\expar}[1]{\mathbb{E}[#1]}

\newcommand{\pr}[1]{\mathrm{Pr}\left[#1\right]}
\newcommand{\prpar}[1]{\mathrm{Pr}[#1]}

\newcommand{\myominus}{\mathbin{\circleddash}}
\newcommand{\probname}{1|\mathrm{rep}|\max_j \sum_i C_{i,j}}
\newcommand{\mylp}{$\mathrm{(LP)}$\xspace}

\sectionfont{\large} \subsectionfont{\normalsize}
\allowdisplaybreaks
\newcounter{dvircounter}
\newcounter{hermelincounter}
\newcounter{segevcounter}

\begin{document}

\begin{titlepage}

\title{Approximation Algorithms for Fair Repetitive Scheduling}
\author{%
Danny Hermelin\thanks{Department of Industrial Engineering and Management, Ben-Gurion University of the Negev, Israel. Email: {\tt \{hermelin,dvirs\}@bgu.ac.il}. Supported by the United States-Israel Binational Science Foundation grant 2024004.}%
\and%
Danny Segev\thanks{School of Mathematical Sciences and Coller School of Management, Tel Aviv University, Tel Aviv 69978, Israel. Email: {\tt segevdanny@tauex.tau.ac.il}. Supported by the Israel Science Foundation grant 1407/20.}%
\and%
Dvir Shabtay\footnotemark[1]}

\date{}
\maketitle

\pagenumbering{Roman}

\begin{abstract}
We consider a recently introduced fair repetitive scheduling problem involving a set of clients, each asking for their associated job to be daily scheduled on a single machine across a finite planning horizon. The goal is to determine a job processing permutation for each day, aiming to minimize the maximum total completion time experienced by any client. This problem is known to be NP-hard for quite restrictive settings, with previous work offering exact solution methods for highly-structured special cases. 

In this paper, we focus on the design of approximation algorithms with provable performance guarantees. Our main contributions can be briefly summarized as follows:
\begin{itemize}
\item When job processing times are day-dependent, we devise a  polynomial-time LP-based $2$-approximation, as well as a polynomial-time approximation scheme for a constant number of days.

\item With day-invariant  processing times, we obtain a surprisingly simple $(\frac{1+\sqrt{2}}{2}+\eps)$-approximation in polynomial time. This setting is also shown to admit a quasi-polynomial-time approximation scheme for an arbitrary number of days.
\end{itemize}
The key technical component driving our approximation schemes is a novel batching technique, where jobs are conceptually grouped into batches, subsequently leading either to a low-dimensional dynamic program or to a compact configuration LP. Concurrently, while developing our constant-factor approximations, we propose a host of lower-bounding mechanisms that may be of broader interest.
\end{abstract}


\end{titlepage}


\newpage
\pagestyle{plain}
\pagenumbering{arabic}
\setcounter{page}{1}

\section{Introduction}
\label{sec:introduction}%

In recent years, a growing body of research has been evolving around  fairness considerations in the design and operation of diverse service-providing systems~\cite{Lipton2004,bertsimas2011fairness,Vardi2025}. This trend is driven by the increasing recognition that such systems, ranging from healthcare and airport slot scheduling to transportation and rent division, can profoundly influence how resources and opportunities are allocated~\cite{Qi17,Breugem2020,Fairbrother2020,Arunachaleswaran2022,Tzur2024}. However, with the rise of algorithmic decision-making and automated resource allocation, these systems may repeatedly favor certain users or groups over others, potentially amplifying existing disparities. Such phenomena have motivated the development of modeling frameworks and algorithmic notions that explicitly incorporate fairness considerations alongside traditional criteria~\cite{bertsimas2012fair,bertsimas2022fairness,Breugem2020,Lipton2004,HebbarSY2024,Babaioff2025}.

Operating within this framework, the current paper focuses on a recently introduced model for injecting fairness features into repetitive scheduling problems~\cite{heeger2021equitable, hermelin2025fairness, shab}. In such problems, we wish to serve a set of $n$ clients across a discrete planning horizon of $m$ days. In the most basic setting, each client requests his individual job to be executed on each day, with a single machine available to process all jobs. Our goal is to determine a set of $m$ schedules, one for each day, that minimize the worst quality of service (QoS) across all clients. Specifically, the latter measure corresponds to the total completion time of each client, one of the most fundamental criteria in scheduling theory. As detailed in Section~\ref{subsec:PreviousWork}, this problem has previously been  studied by Hermelin et al.~\cite{hermelin2025fairness}, who primarily focused on complexity classification and on the design of exact algorithms (either polynomial and parameterized) for highly-structured special cases. Consequently, the approximability of repetitive scheduling in its generality remains largely unexplored.

\subsection{Model description and notation}

\paragraph{Input ingredients and solution concept.} The repetitive scheduling problem we consider includes a set of $n$ clients, as well as a finite planning horizon, consisting of $m$ days. For each day $i \in [m]$, every client $j \in [n]$ has a single job $(i,j)$, with an integer-valued processing time of $p_{i,j}$. On each of these days, all jobs are available at time~0 to be non-preemptively scheduled on a single machine. As such, any feasible solution can be  specified by a sequence $\pi = (\pi_1, \ldots, \pi_m)$ of~$m$ permutations, where each $\pi_i$ represents the processing order of the jobs $\{ (i,j) \}_{j \in [n]}$ on day $i$. For convenience, the terms ``permutations'' and ``schedules'' will be used interchangeably. In addition, $\pi_i(j)$ will denote the position of job $(i,j)$ in the processing sequence $\pi_i$ of day~$i$. Thus, for example, the set $\{\pi^{-1}_i(1), \ldots, \pi^{-1}_i(\ell)\}$ represents the first~$\ell$ jobs scheduled on this day.    

\paragraph{Objective.} With respect to any solution $\pi$, the completion time $C_{i,j}(\pi)$ of job $(i,j)$ is clearly given by  the total processing time of all jobs scheduled before $(i,j)$ on day~$i$, plus its own processing time, i.e., 
\[ C_{i,j}(\pi) ~~=~~ \sum_{ \MyAbove{ \ell \in[n]: }{ \pi_i(\ell) \leq \pi_i(j)}}  p_{i,\ell} \ . \]
With this notation, the total completion time client~$j$ incurs across all $m$ days can be written as $C_j(\pi)= \sum_{i\in[m]} C_{i,j}(\pi)$, with smaller values indicating higher quality of service. Our objective is to compute a solution $\pi$ for the entire planning horizon that minimizes the maximum total completion time $K(\pi)= \max_{j \in [n]} C_j( \pi)$ incurred over all clients. Based on the classical three-field notation, Hermelin et al.~\cite{hermelin2025fairness} refer to this problem by $\probname$.

\subsection{Prior work and open questions}
\label{subsec:PreviousWork}

\paragraph{Hardness results.} As previously mentioned, this  problem has recently been explored by Hermelin et al.~\cite{hermelin2025fairness}, whose first contribution was to establish several hardness results. Specifically, this problem was shown to be weakly NP-hard even for instances with two clients~$(n = 2)$, as well as for instances consisting of four days~$(m = 4)$. Moreover, they proved $\mathrm{W}[1]$-hardness with respect to the number of days~$m$, meaning that an exact $f(m) \cdot n^{O(1)}$-time  algorithm is unlikely to exist, for any computable function $f$. Interestingly, the authors proved that it is NP-hard to distinguish between problem instances for which $K(\pi^*) \leq 37$ and those with $K(\pi^*) \geq 38$, where $\pi^*$   is an optimal schedule. As an immediate byproduct, an approximation factor strictly smaller than $38/37$ cannot be obtained in polynomial time, unless $\mathrm{P} = \mathrm{NP}$.

\paragraph{Algorithmic findings.} On the algorithmic front, a handful of positive results
have been derived for highly-structured special cases. First, Hermelin et al.~\cite{hermelin2025fairness} proposed an exact algorithm for two-day instances, leaving the complexity status of $m=3$ as an open question. Moreover, they designed a fixed-parameter tractable algorithm with respect to the combined parameter $m+\bar{K}$, where $\bar{K}$ is any given upper bound on the optimum value, i.e., the running time obtained is $f(m+\bar{K}) \cdot n^{O(1)}$, for some function~$f$. Finally, the $\probname$ problem was shown to admit an exact $(f(n+\bar{K}))^{O(n)}$-time algorithm, which is pseudo-polynomial for a constant number of clients, i.e., $n=O(1)$. For such instances, this result can be converted into a fully polynomial-time approximation scheme (FPTAS) by standard scaling ideas, where the processing time of each job is scaled down by a factor of $\frac{ \epsilon  \bar{K} }{ mn }$, and then rounded down to the nearest integer.

Building on these results, Plotkin et al.~\cite{shab} proposed several heuristic approaches for addressing $\probname$ and experimentally evaluated their performance. These heuristics include two greedy algorithms, a day-insertion procedure, and a simulated annealing approach. In addition, they studied the special case of unit processing times, and established a lower bound of $\lceil \frac{ (n+1)m }{ 2 } \rceil$ on the objective value of any solution. Then, they proposed a polynomial-time algorithm for computing a solution matching this  bound.

\paragraph{Open questions.} In summary, the results described above largely settle basic questions regarding lower bounds on the approximability of the $\probname$ problem and those related to its parameterized complexity with respect to the number of days $m$ and the number of clients $n$. Nevertheless, quite a few fundamental challenges remain wide open, particularly regarding the approximability of this problem in its utmost generality. We proceed by succinctly shedding light on these challenges from several different angles:
\begin{itemize}
\item {\em Non-trivial approximation guarantees}: Given  the APX-hardness result of Hermelin et al.~\cite{hermelin2025fairness}, the first challenge is that of attaining a constant-factor approximation for an arbitrary number of clients and days. Yet another question along these lines is whether one can design an approximation scheme for a constant number of days, i.e., when $m=O(1)$.

\item {\em Developing lower bounds}: A central obstacle toward resolving these questions resides in identifying  effective upper and lower bounds on the optimum value. While several bounds in this context were proposed by earlier papers, none appears to be sufficiently tight to provide constant-factor approximation, let along approximation schemes. 

\item {\em Improved guarantees for identical days}: Finally, the setting of day-invariant processing times has not been studied prior to our work, despite its relevance in real-world scenarios. Here, a fundamental question is whether this setting is ``easier'' to approximate in comparison to its day-dependent counterpart.
\end{itemize}

\subsection{Main results}

The main contributions of this paper reside in resolving all  challenges discussed above. In particular, for the most general setting of day-dependent processing times, we develop a constant-factor approximation as well as a polynomial-time approximation scheme for a constant number of days. Subsequently, we provide improved approximation guarantees for day-invariant processing times, establishing several novel lower bounds along the way. In what follows, we discuss our results in greater detail, touching upon selected technical ideas. For ease of exposition, these two versions of the $\probname$ problem will be referred to as the day-dependent model and the day-invariant model, respectively.

\paragraph{The day-dependent model: $\boldsymbol{2}$-approximation.} In Section~\ref{sec:2_approx_general}, we consider the most general setting, where the processing time incurred by each client's job may vary across different days. In this context, we design a polynomial-time $2$-approximation, based on LP-rounding. Interestingly, the relaxation we utilize is inspired by that of Hall et al.~\cite{HallEtAl1997}, who developed a $3$-approximation for minimizing the sum of completion times on a single machine with arbitrary release dates. While our relaxation is comprised of exponentially-many constraints, it is polynomial-time solvable using the Ellipsoid method, via an appropriate separation oracle. Then, given an optimal fractional solution, we construct a feasible schedule whose objective value is within factor $2$ of our LP-based bound. 

\begin{theorem}
\label{thm:2Approx}%
The day-dependent model can be approximated in polynomial time within factor $2$ of optimal.
\end{theorem}

\paragraph{The day-dependent model: PTAS for $\boldsymbol{m=O(1)}$ days.} Our second result, whose specifics are provided in Section~\ref{sec:PTAS}, comes in the form of a polynomial-time approximation scheme (PTAS) for day-dependent instances consisting of $m=O(1)$ days. Our PTAS relies on a newly-devised batching scheme, where each batch represents a set of jobs that are grouped together and treated as being simultaneously processed. In particular, we define a structured set of ``good'' batchings, that can be efficiently enumerated. For each such batching, we employ dynamic programming ideas to compute a $(1+O(\eps))$-approximate job-to-batch assignment,  leading to the next result.\footnote{Here, the $\tilde{O}$ notation suppresses polylogarithmic factors, meaning that $\tilde{O}(f(n))=O(f(n) \cdot \log^{O(1)} f(n))$.} 

\begin{theorem}
\label{thm:PTAS}%
For any $\eps > 0$, the day-dependent model can be approximated within factor $1+\eps$ of optimal. Our algorithm can be implemented in $O(n^{\tilde{O}(m / \eps^2)})$ time.
\end{theorem}

\paragraph{The day-invariant model: Improved constant-factor approximation.} In Section~\ref{sec:identical_better}, we focus our attention on day-invariant instances, where each client incurs a uniform processing time across all~$m$ days. In other words, $p_{i,j}=p_j$ for every client $j \in [n]$ and day $i \in [m]$. In this context, we show that the uniformity in processing times across different days allows us to uncover additional structural properties, leading to an improvement on the $2$-approximation obtained for day-dependent instances (see Theorem~\ref{thm:2Approx}). Specifically, we establish a novel lower bound on the objective value of any solution, which is exploited to argue that a rather simple schedule achieves a significantly better approximation. This result can be formally stated as follows.

\begin{theorem} \label{thm:BetterApprox}%
For any $\epsilon > 0$, the day-invariant model can be approximated in $O(n^{\tilde{O}(1/\eps^3)})$ time within factor $\frac{1+\sqrt{2}}{2}+\epsilon$ of optimal.
\end{theorem}

\paragraph{The day-invariant model: Quasi-PTAS.} Lastly, in Section~\ref{sec:QPTAS}, we present a quasi-polynomial-time approximation scheme (QPTAS) for day-invariant instances, defined with respect to an arbitrary number of days. To this end, we first show that to obtain a $(1+\eps)$-approximation, it suffices to  focus on instances consisting of $m=O(\frac{\log n}{\eps^3})$ days. Then, we refine the notion of good batchings, tailoring it to the simpler structure of day-invariant processing times. Finally, rather than resorting to a dynamic programming approach, as in Section~\ref{sec:PTAS}, we formulate a configuration LP, and make use of randomized rounding to compute a near-optimal job-to-batch assignment. As formally stated below, these ideas culminate to an approximation scheme sitting at the very low end of quasi-polynomial time, with an exponential dependency on $\log\log n$.

\begin{theorem} \label{thm:QPTAS}
For any $\eps > 0$, the day-invariant model can be approximated within factor $1+\eps$ of optimal. Our algorithm can be implemented in $O(n^{\tilde{O}( \frac{ \log \log n }{ \eps^9 })})$ time. 
\end{theorem}

\subsection{Related work}

In what follows, we briefly survey three directly-related research streams, with the objective of providing additional background to avid readers. For this purpose, we first review studies on fairness in repetitive scheduling, focusing primarily on alternative quality of service measures rather than on total completion time. Then, we discuss the usefulness of configuration LPs in the design of approximation algorithms. Finally, we elaborate on selected algorithmic results for minimizing total completion time in standard (non-repetitive) scheduling problems.

\paragraph{Fairness in repetitive scheduling.} The framework of repetitive scheduling studies scenarios in which each client submits a job to be processed on each day of a given planning horizon, and performance is measured in terms of client-specific service levels across all days. The first work in this framework is that of Heegar et al.~\cite{heeger2021equitable}, who considered the problem of minimizing the maximum number of tardy jobs per client over all days, known as $1|\mathrm{rep}|\max_j \sum_i U_{i,j}$. Here, $U_{i,j}$ is a binary variable, indicating whether job $(i,j)$ completes after its due date. The authors established hardness results for several restricted variants, along with algorithmic results for special cases.

Later on, Hermelin et al.~\cite{hermelin2025fairness} extended the repetitive scheduling framework to additional objective functions. Apart from the maximum total completion time objective, which is thoroughly discussed in Section~\ref{subsec:PreviousWork}, they studied minimizing the maximum total waiting time as well as minimizing the maximum total lateness of any client across all days. For each of these objectives, the authors obtained results of a similar nature, combining hardness classifications with algorithmic contributions under restricted settings.

More recently, Heegar et al.~\cite{heeger2025interval} analyzed the just-in-time criterion within the repetitive scheduling framework. In this setting, each job $(i,j)$ can either be processed within a predefined time interval $(d_{i,j}-p_{i,j},d_{i,j}]$ or not processed at all (i.e., rejected). Here, the time intervals for processing different jobs cannot intersect, with quality of service measured by the number of scheduled jobs for each client. Heegar et al.~\cite{heeger2025interval} provided complexity classifications and efficient algorithms for the just-in-time criterion in the repetitive scheduling scenario.

\paragraph{Configuration LPs.} Configuration linear programs are powerful relaxations, widely used in the design and analysis of approximation algorithms for various combinatorial optimization problems. Unlike standard LP formulations, which typically include a variable for singular decisions, configuration LPs introduce a variable for every feasible subset (or configuration) of decisions to be jointly made, thereby capturing the combinatorial structure of a given problem in greater granularity. While this approach often leads to exponentially-many variables, configuration LPs can be efficiently solved/approximated in certain cases via techniques such as column generation, and their fractional solutions provide strong bounds that guide LP-rounding algorithms. Interestingly, they have played a central role in obtaining state-of-the-art approximation guarantees for a wide range of problems, including bin packing \cite{Karmarkar82,Bansal2006,Kulik2023}, makespan minimization on unrelated parallel machines \cite{Lenstra90, Svensson2012, VerschaeW14, KnopK22}, scheduling to minimize total completion time with release dates \cite{Phillips,Chekuri2001}, and  capacitated facility location  \cite{chekuri2004}.

\paragraph{Approximations for  completion time minimization.} Minimizing total completion time is one of the most classical scheduling objectives \cite{Pinedo12scheduling, ChekuriK04, AgnetisBPS25}, with an immense body of work on its standard (non-repetitive) setting. When all jobs are available at time $0$, the single-machine problem can be solved in $O(n\log n)$ time via the well-known SPT rule~\cite{Smith1956}. However, when jobs are associated with  arbitrary release dates, this problem becomes strongly NP-hard~\cite{LRKB1977}. Nevertheless, approximation results in this setting (denoted by $1|r_j|\sum_j C_j$) have been an extensively studied topic, and we elaborate on selected contributions below. 

Phillips et al.~\cite{Phillips} were the first to provide a constant-factor approximation for the $1|r_j|\sum_j C_j$ problem. The main idea behind their approach is a simple method for converting a preemptive schedule into a non-preemptive one, increasing the completion time of each job by a factor of at most $2$. Since an optimal preemptive schedule can be efficiently computed, this relation leads to a polynomial-time $2$-approximation. Chekuri et al.~\cite{Chekuri} enhanced the latter approach by developing randomized algorithms that attain an approximation ratio of $\frac{ e }{ e-1 } \approx 1.582$ for the $1|r_j|\sum_j C_j$ problem, and a $2.83$-approximation for its extension, $\mathrm{P}|r_j|\sum_j C_j$, where jobs are scheduled on an arbitrary number of identical parallel machines. Here, given an $m$-machine instance, their main idea is to propose a single-machine preemptive relaxation, and then to make use of its optimal solution to design a feasible schedule  for the original $m$-machine instance. 

Hall et al.~\cite{HallEtAl1997} provided a $3$-approximation for the weighted sum of completion time objective, designated by $1|r_j|\sum_j w_jC_j$. To this end, they exploited an LP relaxation with exponentially-many constraints, which is still solvable in polynomial time. Then, adopting the approach of Phillips et al.~\cite{Phillips}, they construct a $3$-approximate schedule out of an optimal fractional solution. Later on, Goemans et al.~\cite{Goemans} attained an improved $\frac{ e }{ e-1 }$-approximation, whereas Skutella~\cite{DBLP:journals/jacm/Skutella01} proposed a $2$-approximation for unrelated machines, based on convex quadratic and semidefinite programming relaxations. Finally, Afrati et al.~\cite{Foto1999} presented polynomial-time approximation schemes for several different  settings of the $\mathrm{P}|r_j|\sum_j C_j$ problem. For additional background on approximation algorithms for minimizing total completion time, we refer interested readers to the excellent survey of Chakuri and Khanna~\cite{ChekuriK04}. 
\section{The Day-Dependent Model: \texorpdfstring{$\boldsymbol{2}$}{}-Approximation} \label{sec:2_approx_general}

In what follows, we show that the day-dependent model can be approximated in polynomial time within factor $2$ of optimal, as formally stated in Theorem~\ref{thm:2Approx}. 
Our approach is based on a new linear relaxation, inspired by the $3$-approximation of Hall et al.~\cite{HallEtAl1997} for minimizing weighted completion times on a single machine, which itself draws on polyhedral results due to Queyranne~\cite{Queyranne1993}. Moving forward, in Sections~\ref{subsec:day_dep_relax}
and~\ref{subsec:Oracle}, we  describe our LP relaxation and explain why it can be solved in polynomial time. Then, in Section~\ref{subsec:day_dep_twoapp}, we show how an optimal fractional solution can be converted into a $2$-approximate schedule.

\subsection{LP relaxation} \label{subsec:day_dep_relax}

\paragraph{The linear relaxation.} Our relaxation introduces a continuous decision variable $x_{i,j}$ for every day~$i \in [m]$ and client $j \in [n]$, representing the completion time of job $(i,j)$. In addition, the  auxiliary continuous variable $K$ will represent the maximal sum of completion times across all clients. With these conventions, the LP relaxation we consider can be compactly written as follows:
\begin{alignat}{3}
& \text{min} &\quad & K \notag \\
&\text{s.t.} &\quad & \sum_{i \in [m]} x_{i,j} \leq K & \qquad \qquad & \forall \, j \in [n] \label{LP:const_1} \\
& &\quad & \sum_{j\in S} p_{i,j}x_{i,j} \geq \frac{1}{2} \cdot (P_{i}(S))^2  & \qquad \qquad & \forall \, i \in [m], \, S\subseteq [n] \label{LP:const_2} 
\end{alignat}
This program consists of $O(nm)$ variables and $O(2^n m)$ constraints. Here, constraint~\eqref{LP:const_1} ensures that $K$ is an upper bound on total completion time of every client $j \in [n]$. Then, for every day $i \in [m]$ and subset $S \subseteq [n]$ of clients, constraint~\eqref{LP:const_2} imposes a lower bound of $\frac{1}{2} \cdot (P_{i}(S))^2$ on the $p_{i,\cdot}$-weighted completion times $\sum_{j\in S} p_{i,j}x_{i,j}$ of the jobs belonging to clients in $S$. In the former expression, $P_i(S)=\sum_{j \in S} p_{i,j}$ denotes the  total processing time of these jobs.

\paragraph{The LP-optimum forms a lower bound.} To verify that this program indeed constitutes a lower bound on the objective value $K(\pi)$ of any schedule $\pi$, let us construct a candidate LP-solution out of $\pi$ by setting $\hat{K}=K(\pi)$ and $\hat{x}_{i,j}=C_{i,j}(\pi)$ for all $i\in[m]$ and $j\in[n]$. Then, constraint~\eqref{LP:const_1} is clearly satisfied, since 
\[ \hat{K} ~~=~~ K(\pi) ~~=~~ \max_{j \in [n]} \sum_{i \in [m]} C_{i,j}(\pi) ~~=~~ \max_{j \in [n]} \sum_{i \in [m]} \hat{x}_{i,j} \ . \]
To see why constraint~\eqref{LP:const_2} is satisfied as well, consider some day~$i \in [m]$ and some subset of clients $S \subseteq [n]$.  Clearly, $(\hat{K},\hat{x})$ satisfies constraint~\eqref{LP:const_2} when $|S|=1$, since assuming that $S = \{j \}$,
\[ p_{i,j} \hat{x}_{i,j} ~~=~~ p_{i,j}  C_{i,j}(\pi) ~~\geq~~ p_{i,j}^2 ~~=~~ (P_i(S))^2 \ . \]
Now, consider some subset of clients $S$ with $|S| \geq 2$. For every day $i \in [m]$, the sum $\sum_{j\in S} p_{i,j}C_{i,j}(\pi)$ includes the term $p_{i,j}p_{i,k}$ for any pair of (not necessarily distinct) clients $j,k\in S$, as either the job of client~$j$ is scheduled by $\pi_i$ before that of client~$k$, or vise versa. Accordingly, we are indeed meeting constraint~\eqref{LP:const_2}, since 
\begin{eqnarray*}
\sum_{j\in S} p_{i,j} \hat{x}_{i,j} & = & \sum_{j\in S} p_{i,j} C_{i,j}(\pi) \\
&\geq & \sum_{ \MyAbove{ j,k \in S : }{ j\leq k} }p_{i,j}p_{i,k} \\
& = & \frac{1}{2} \cdot \left(\sum_{j\in S}p_{i,j} \right)^{2}+ 
\frac{1}{2}\sum_{j\in S}p^2_{i,j} \\ 
& > & \frac{1}{2} \cdot (P_{i}(S))^2.
\end{eqnarray*}
Thus, $(\hat{K},\hat{x})$ is indeed a feasible solution to our relaxation, with an objective value of $\hat{K}=K(\pi)$.

\subsection{Polynomial-time separation oracle}
\label{subsec:Oracle}%

As previously mentioned, our LP relaxation has an exponential number of constraints, due to $O(2^n m)$ constraints of type~\eqref{LP:const_2}. Nevertheless, this LP can still be solved in polynomial time via the Ellipsoid method, assuming we can implement a polynomial-time separation oracle~\cite{GrotschelEtAl81}. Such a procedure receives a vector $(K,x)$ as input, and decides whether it is a feasible solution or not. In the latter case, it finds a violated constraint, which is either a client $j\in[n]$ with $\sum_{i \in [n]} x_{i,j} > K$, or a subset of clients $S \subseteq [n]$ with $\sum_{j\in S} p_{i,j}x_{i,j} < \frac{1}{2} \cdot (P_{i}(S))^2$ for some day $i \in [m]$. An algorithm for the first type of violation is trivial, due to having only $n$ clients to examine; for the second type, we proceed by elaborating on two possible options.

\paragraph{Option 1: Submodularity.} First, focusing on a single day $i \in [m]$, it is not difficult to verify that the function $f_i( S ) = \sum_{j \in S} p_{i,j} x_{i,j} - \frac{1}{2} \cdot (P_i(S))^2$ is submodular. Since the latter expression is precisely constraint~\eqref{LP:const_2} in rearranged form, it is satisfied for every subset $S \subseteq [n]$ if and only if $\min_{S \subseteq [n]} f_i(S) \geq 0$. As such, we are left with solving a submodular minimization problem, for which there are several well-known polynomial-time algorithms (see, for example, \cite{IwataFF01, Schrijver00, Orlin09}). 

\paragraph{Option 2: A direct approach.} Alternatively, following arguments similar to those of Queyranne~\cite{Queyranne1993}, we propose a more efficient method in comparison to the submodularity-based oracle, tailor made to the specific structure of constraint~\eqref{LP:const_2}. Our separation oracle iterates over all days $i \in [m]$, and for each day~$i$ proceeds as follows: First, it computes an ordering of the clients, $\pi_i$, by non-decreasing value of $x_{i,j}$. That is, $\pi_i$ is a permutation of the jobs such that $\pi_i(j) < \pi_i(\ell)$ whenever $x_{i,j} < x_{i,\ell}$. Next, for each $\ell \in [n]$, it considers the subset 
$S_\ell^{(i)}=\{\pi^{-1}_i(1),\ldots,\pi^{-1}_i(\ell)\}$,
and checks whether it violates constraint~\eqref{LP:const_2}; namely, whether $\sum_{j\in S_{\ell}^{(i)}} p_{i,j}x_{i,j} < \frac{1}{2} \cdot (P_{i}(S_\ell^{(i)}))^2$. When such a violation is detected, we have just identified a constraint that is violated by $(K,x)$. In the opposite case, we proceed to day $i+1$. Clearly, the entire procedure can be implemented in $O(mn \log n)$ time. 

The next lemma shows that restricting our attention to $S_1^{(i)}, \ldots, S_n^{(i)}$ is sufficient, arguing that when $(K,x)$ is infeasible, a violated constraint will always be detected. 

\begin{lemma}
Suppose that $(K,x)$ satisfies constraint~\eqref{LP:const_2} for each of the subsets $S_1^{(i)}, \ldots, S_n^{(i)}$. Then, it is satisfied for any subset $S \subseteq [n]$ on day $i$.      
\end{lemma}
\begin{proof}
Suppose that there exists a subset of clients $S \subseteq [n]$ for which $\sum_{j\in S} p_{i,j}x_{i,j} < \frac{1}{2} \cdot (P_{i}(S))^2$, corresponding to a violation of constraint~\eqref{LP:const_2} on day~$i$. We show that there exists some $\ell \in[n]$ for which $\sum_{j\in S_{\ell}^{(i)}} p_{i,j}x_{i,j} < \frac{1}{2} \cdot (P_{i}(S_\ell^{(i)}))^2$. To this end, consider the following two-step procedure:
\begin{enumerate}
\item Let $\hat{\ell}$ be the maximal-index element in $S$. If $x_{i,\hat{\ell}} > P_i(S \setminus \{\hat{\ell}\})+\frac{1}{2} p_{i,\hat{\ell}}$, then remove $\hat{\ell}$ from $S$ and repeat this step. Otherwise, proceed to step~2.

\item Let $\hat{\ell}$ be the maximal-index element in $S$. If there exists a client $k \in S_{\hat{\ell}}^{(i)} \setminus S$, then add $k$ to $S$ and repeat this step. Otherwise, terminate.
\end{enumerate}

This procedure clearly terminates with $S=S_\ell^{(i)}$ for some $\ell \in [n]$. We claim that $(K,x)$ violates constraint~\eqref{LP:const_2} with respect to $S_\ell^{(i)}$ on day~$i$. For this purpose, let 
\[ \Delta_i(S) ~~=~~\frac{1}{2} \cdot (P_{i}(S))^2 - \sum_{j\in S} p_{i,j}x_{i,j} ~~>~~ 0 \] 
be the difference between the right-hand and left-hand sides of  constraint~\eqref{LP:const_2}, corresponding to~$S$ on day~$i$. We show that $\Delta_i(\cdot)$ can only increase throughout the procedure described in steps~1 and~2. Indeed, regarding step~1, when client $\hat{\ell}$ is removed, we have
\begin{eqnarray*}
\Delta_i(S) &=& \frac{1}{2} \cdot (P_{i}(S))^2 - \sum_{j\in S} p_{i,j}x_{i,j} \\
&=& \frac{1}{2} \cdot (P_{i}(S\setminus \{\hat{\ell}\}))^2 + p_{i,\hat{\ell}} \cdot P_i(S \setminus \{\hat{\ell}\}) +  \frac{1}{2} p^2_{i,\hat{\ell}}  - \sum_{j\in S} p_{i,j}x_{i,j} \\
&< & \frac{1}{2} \cdot (P_{i}(S\setminus \{\hat{\ell}\}))^2 + p_{i,\hat{\ell}}x_{i,\hat{\ell}}  - \sum_{j\in S} p_{i,j}x_{i,j} \\
&=& \frac{1}{2} \cdot P^2_{i}(S \setminus \{\hat{\ell}\}) - \sum_{j\in S \setminus \{\hat{\ell}\}} p_{i,j}x_{i,j} \\
& = & \Delta_i(S\setminus\{\hat{\ell}\}),
\end{eqnarray*}
where the inequality above holds since $x_{i,\hat{\ell}} > P_i(S \setminus \{\hat{\ell}\}) +\frac{1}{2} p_{i,\hat{\ell}}$. Accordingly, we get $\Delta_i(S) < \Delta_i(S\setminus\{\hat{\ell}\})$. Similarly, in step~2, for any job $k$ that is added to the set $S$, we have $x_{i,k} \leq x_{i,\hat{\ell}} \leq P_i(S) < P_i(S)+\frac{1}{2}p_{i,k}$, implying that $p_{i,k}x_{i,k} \leq p_{i,k} \cdot P_i(S)+\frac{1}{2}p^2_{i,k}$. Thus, 
\begin{eqnarray*}
\Delta_i(S) &=& \frac{1}{2} \cdot (P_{i}(S))^2 - \sum_{j\in S} p_{i,j}x_{i,j} \\
&=& \frac{1}{2} \cdot (P_{i}(S\cup \{k\}))^2 - p_{i,k} \cdot P_i(S) -  \frac{1}{2}p^2_{i,k}  - \sum_{j\in S} p_{i,j}x_{i,j} \\
&<& \frac{1}{2} \cdot (P_{i}(S\cup \{k\}))^2 - p_{i,k}x_{i,k}  - \sum_{j\in S} p_{i,j}x_{i,j} \\
&=&\frac{1}{2} \cdot (P_{i}(S \cup \{k\}))^2 - \sum_{j\in S \cup \{k\}} p_{i,j}x_{i,j} \\
& = & \Delta_i(S\cup\{k\}),
\end{eqnarray*}
and once again, $\Delta_i(S)<\Delta_i(S\cup \{k\})$. 
\end{proof}

\subsection{Approximate schedule and analysis}
\label{subsec:day_dep_twoapp}%

\paragraph{Converting fractional solutions to schedules.} Let $(K^*,x^*)$ be an optimal fractional solution to our LP relaxation, computed in polynomial time as discussed above. We utilize this solution to construct an approximate schedule $\pi=(\pi_1,\ldots,\pi_m)$ as follows. For each day $i \in [m]$, its schedule $\pi_i$ will be a permutation of the clients in non-decreasing order of their $x^*_{i,\cdot}$-value, namely, $x^*_{i,\pi^{-1}_i(1)}\leq x^*_{i,\pi^{-1}_i(2)},\ldots,\leq x^*_{i,\pi^{-1}_i(n)}$. In this way, the completion time of each job $(i,j)$ is given by $C_{i,j}( \pi) =\sum_{k \in [n]: \pi_i(k) \leq \pi_i(j)} p_{i,k}$.

\paragraph{Approximation guarantee.} To show that our  schedule $\pi$ forms a $2$-approximation, we first prove that $C_{i,j}( \pi ) \leq 2x^*_{i,j}$, for every day  $i\in[m]$ and client $j\in[n]$. For this purpose, let $\ell=\pi^{-1}_i(j)$ be the position of the job $(i,j)$ with respect to $\pi_i$. Now, consider the subset of clients $S_\ell^{(i)} = \{\pi^{-1}_i(1), \ldots, \pi^{-1}_i(\ell)\}$, i.e., those  whose jobs are scheduled no later than $(i,j)$ on day~$i$, including the job $(i,j)$ itself. Then, according to our construction,  $C_{i,j}( \pi ) =P_i(S_\ell^{(i)})$. On the other hand, since $(K^*,x^*)$ is a feasible LP solution, it satisfies constraint~\eqref{LP:const_2} for $S_\ell^{(i)}$ on day $i$, implying that
\[ \frac{1}{2} \cdot (P_i(S_\ell^{(i)}))^2 ~~\leq~~ \sum_{k \in S_\ell^{(i)}} p_{i,k} x^*_{i,k}  ~~\leq~~ x^*_{i,j} \cdot \sum_{k\in S_\ell^{(i)}} p_{i,k} ~~=~~ x^*_{i,j} \cdot P_{i}(S_\ell^{(i)}), \]
where the second inequality holds since $x^*_{i,j} \geq x^*_{i,k}$ for all $k \in S_\ell^{(i)}$, by definition of $S_\ell^{(i)}$. By rearranging, it follows that $2x^*_{i,j} \geq P_{i}(S_\ell^{(i)}) = C_{i,j}( \pi )$. 

Now, to conclude that $\pi$ is a $2$-approximate schedule, we observe that every client $j\in[n]$ has a total completion time of 
\[ C_j(\pi) ~~=~~ \sum_{i\in[m]} C_{i,j}(\pi) ~~\leq~~ 2 \cdot \sum_{i\in[m]} x^*_{i,j} ~~\leq~~ 2K^*, \]
where the last inequality holds since $\sum_{i\in[n]} x^*_{i,j} \leq K^*$, due to constraint~\eqref{LP:const_1}.
\section{The Day-Dependent Model: PTAS for \texorpdfstring{$\boldsymbol{m=O(1)}$}{} Days}
\label{sec:PTAS}

This section is dedicated to proving that, for a constant number of days, the day-dependent model admits a polynomial-time approximation scheme. Specifically, as formally stated in Theorem~\ref{thm:PTAS}, we will show that this model can be approximated within factor $1+\eps$ of optimal in $O(n^{\tilde{O}(m/\eps^2)})$ time.

\paragraph{Outline.} The main technical idea leading to the above-mentioned result is the notion of batchings and their associated assignments, which are introduced in Section~\ref{subsec:BatchAssign}. Intuitively, each batch represents a set of jobs that are grouped together, and subsequently thought of as being simultaneously processed. In Sections~\ref{subsec:GoodBatchings} and~\ref{subsec:StructureLemmaProof}, we will focus on a structured subclass of ``good'' batchings, proving that they preserve sufficiently fine-grained information to approximate the optimal schedule within any degree of accuracy. Additionally, due to having polynomially-many such batchings, they can be exhaustively enumerated. Finally, in Section~\ref{subsec:PTAS_DP}, we present a dynamic programming procedure that, given a good batching, computes an approximate  job-to-batch assignment.

\subsection{Batchings and assignments} 
\label{subsec:BatchAssign}%

\paragraph{Definitions and notation.} A batching for day $i \in [m]$ is defined as a sequence $B_i = (B_{i,1}, \ldots, B_{i,\beta})$ of non-negative integers, where each $B_{i,b}$ denotes the capacity of the $b$-th batch, also referred to as batch $(i,b)$. The number of batches, $\beta$, is called the length of this batching. Extending this definition to all days, we say that $B=(B_1, \ldots, B_m)$ is a batching of length~$\beta$  when each $B_i$ is a batching for day~$i$ of length $\beta$. Here, the latter length is identical for all days, whereas the capacities $B_{i,b}$ may vary both across batches within a single day and across different days.

Given such a batching $B_i$ for day~$i$, we refer to any function $A_i : [n] \to [\beta]$ as a job-to-batch assignment. In this way, $A_i(j)$ denotes the batch job~$(i,j)$ is assigned to, and conversely, $A_i^{-1}(b)$ denotes the set of jobs assigned to batch~$(i,b)$. An assignment $A_i$ is called feasible if, for every $b \in [\beta]$, the total processing time of all jobs assigned to batch $(i,b)$ fit within its capacity, i.e., $\sum_{j \in A_i^{-1}(b)} p_{i,j} \leq B_{i,b}$. When this capacity is exceeded by an additive factor of at most $\Delta$, namely, $\sum_{j \in A_i^{-1}(b)} p_{i,j} \leq B_{i,b} + \Delta$, the assignment $A_i$ is said to be $\Delta$-feasible. These definitions naturally extend to an overall assignment, with $A = (A_1, \ldots, A_m)$ being  ($\Delta$-)feasible with respect to a batching $B=(B_1,\ldots,B_m)$ when $A_i$ is \mbox{($\Delta$-)feasible} with respect to $B_i$, for every day $i \in [m]$.

\paragraph{Batching-related objective function.} When jobs are assigned to a batch, they will be viewed as being simultaneously processed and completed at the same time. To formalize this concept, we define the end ${\cal E}^B_{i,b}$ of batch $(i,b)$ by ${\cal E}^B_{i,b} = \sum_{\hat{b} \leq b} B_{i,\hat{b}}$, which is the total capacity of all batches up to and including $(i,b)$. In turn, for a given batching $B = (B_1, \ldots, B_m)$ and a corresponding assignment $A = (A_1, \ldots, A_m)$, we observe that $\sum_{i \in [m]} {\cal E}^B_{i,A_i(j)}$ represents the total ends of the batches to which all jobs of client $j$ are assigned. By taking the maximum of this measure over all clients, we create a new objective function, operating on batching-assignment pairs:
\[ K(A,B) ~~=~~ \max_{j \in [n]} \sum_{i \in [m]} {\cal E}^B_{i,A_i(j)}. \]

\paragraph{Translating batching-assignment pairs to schedules.} In what follows, we prove that any $\Delta$-feasible assignment $A$ with respect to any batching $B$ can be efficiently transformed into a schedule $\pi$, while ensuring that $K(\pi)$ and $K(A,B)$ are not too far off. 

\begin{lemma} 
\label{lem:AssignmentToSchedule}%
Let $B$ be a batching of length $\beta$, and let $A$ be a $\Delta$-feasible assignment with respect to~$B$. Then, $A$ can be translated in $O(nm)$ time to a schedule $\pi$ such that 
\[ K(\pi) ~~\leq~~  K(A,B) + m \beta\Delta. \]
\end{lemma}
\begin{proof}
To construct a schedule $\pi_i$ for each day $i \in [m]$, we process the set of jobs $A_i^{-1}(b)$ assigned to each batch $(i,b)$ in an arbitrary order. The resulting permutations are then glued together by increasing batch index, i.e., the $A_i^{-1}(1)$-ordering appears first, followed by the $A_i^{-1}(2)$-ordering, so on and so forth. Since the sets $\{A_i^{-1}(b)\}_{b \in [\beta]}$ form a partition of all jobs on day~$i$, this procedure ensures that $\pi_i$ is indeed a complete schedule for day $i$. With respect to the resulting solution $\pi=(\pi_1,\ldots,\pi_m)$, the completion time $C_{i,j}(\pi)$ of each job $(i,j)$ can be upper-bounded by observing that
\begin{eqnarray*} 
C_{i,j}(\pi) &\leq & \sum_{ \MyAbove{k \in [n]:}{ A_i(k) \leq A_i(j)}} p_{i,k} \\
& = & \sum_{b \leq A_i(j)} \sum_{ \MyAbove{k \in [n]:}{ A_i(k) = b}} p_{i,k} \\
&\leq & \sum_{b \leq A_i(j)} (B_{i,b} + \Delta) \\
& \leq & {\cal E}^B_{i,A_i(j)} + \beta \Delta,
\end{eqnarray*}
where the middle inequality holds since $A$ is a $\Delta$-feasible assignment with respect to $B$. By summing the above inequality across all days, we conclude that the total completion time any client $j\in[n]$ incurs is 
\[ C_j(\pi) ~~=~~ \sum_{i \in [m]} C_{i,j}(\pi) ~~\leq~~ \sum_{i \in [m]} {\cal E}^B_{i,A_i(j)} + m \beta \Delta, \]
and therefore,
\[ K(\pi) ~~=~~ \max_{j \in [n]} C_j(\pi) ~~\leq~~ \max_{j \in [n]} \sum_{i \in [m]} {\cal E}^B_{i,A_i(j)} + m \beta \Delta ~~=~~  K(A,B) + m \beta\Delta.\]
\end{proof}

\paragraph{Setting $\boldsymbol{\Delta}$.} For the remainder of this section, we choose the value of $\Delta$ to be sufficiently small so that the $m\beta\Delta$-term in Lemma~\ref{lem:AssignmentToSchedule} becomes negligible. As shown in Section~\ref{subsec:GoodBatchings}, it suffices to restrict our attention to batchings of length $\beta = O (\frac{ \log m }{\eps^2} )$. Thus, to ensure that $m\beta\Delta$ is negligible, we set
\begin{equation}
\label{eqn:Delta}%
\Delta ~~=~~ \frac{\eps^3}{m^2} \cdot \tilde{K}.
\end{equation}
Here, $\tilde{K}$ is a $(1+\eps)$-estimate for the objective value $K^* = K( \pi^* )$ of an optimal schedule $\pi^*$, meaning that $K^* \leq \tilde{K} \leq (1+\eps) K^*$. Recalling that Theorem~\ref{thm:2Approx} provides a constant-factor approximation for $K^*$, standard arguments show that such an estimate can be obtained by testing $O( \frac{ 1 }{ \eps } )$ integer powers of $1+\eps$ as candidate values. 

\subsection{Good batchings} 
\label{subsec:GoodBatchings}%

\paragraph{Preliminaries.} Informally, we refer to a batching $B=(B_1,\ldots,B_m)$ of length~$\beta$ as being  ``good'' from an algorithmic perspective when both~$\beta$ and the number $|\{B_{i,b} : i \in [m], b \in [\beta]\}|$ of distinct batch capacities are relatively small. To quantify this intuition, let us begin by defining the discrete set of values
$$
{\cal B} ~~=~~ \left\{ \Delta,(1 + \eps) \cdot \Delta, (1 + \eps)^2 \cdot \Delta, \ldots, (1 + \eps)^\chi \cdot \Delta   \right\}, 
$$
where $\chi$ is the smallest integer for which $(1 + \eps)^\chi \cdot \Delta \geq \tilde{K}$, which is equivalent to $(1 + \eps)^\chi \geq \frac{ m^2 }{ \eps^3 }$, by recalling that $\Delta = \frac{\eps^3}{m^2} \cdot \tilde{K}$. This choice implies that 
\begin{eqnarray}
\label{eqn:GoodBatchingLength}    
|\mathcal{B}| & = & \left\lceil \log_{1+\eps} \Big(\frac{ m^2 }{ \eps^3 }\Big) \right\rceil \leq  \frac{ 2 \ln m +  3 \ln \frac{ 1 }{ \eps } }{ \ln(1+\eps) } + 1 \nonumber \\
& \leq &  \frac{ 2 \ln m +  3 \ln \frac{ 1 }{ \eps } }{ \eps/2 } + 1   ~~\leq~~ \frac{ 4 \ln m }{ \eps } + \frac{ 6  }{ \eps^2 } + 1  ~~\leq~~ \frac{ 11 \ln m }{ \eps^2 }, 
\end{eqnarray}
where the second inequality holds since $1+\eps \geq e^{\eps/2}$ for $\eps < 1$.

In turn, we define a batching $B=(B_1,\ldots,B_m)$ of length $\beta$ as being good when it satisfies the next two properties:
\begin{enumerate}
\item {\em Short length}: $\beta \leq 2 \cdot |{\cal B}|$.
\item {\em ${\cal B}$-valued capacities}: $B_{i,b} \in {\cal B}$ for every $i \in [m]$ and $b \in [\beta]$.
\end{enumerate}
As such, since $|\mathcal{B}|=O(\frac{\log m }{ \eps^2})$, the number of good batchings is only $O(|{\cal B}|^{\beta \cdot m}) = O( |{\cal B}|^{ 2|{\cal B}| \cdot m } ) = 2^{\tilde{O}(m/\eps^2)}$. 

\paragraph{The structure lemma.} Now consider some optimal schedule $\pi^*=(\pi^*_1,\ldots,\pi^*_m)$ for the entire planning horizon of $m$ days. A straightforward observation is that, with $n$ daily batches and at most $mn$ distinct batch capacities, there exists a batching $B$ and a corresponding feasible assignment $A$ whose objective value $K(A,B)$ is exactly $K^*=K(\pi^*)$. Indeed, one simply creates a separate batch for each job~$j$ in any given day~$i$,  ordered according to $\pi^*_i$, with capacity $p_{i,j}$. Quite surprisingly, our main structural result shows that, even when restricting ourselves to good batchings, which only have $O(\frac{\log m }{ \eps^2})$ distinct batch capacities and length $O( \frac{ \log m }{ \eps^2 })$, this relationship between $K(A,B)$ and~$K^*$ can still be approximately preserved. 

\begin{lemma}[\textbf{Structure Lemma}] 
\label{lem:StructureLemma}
There exists a good batching $B^*$, together with a feasible assignment~$A^*$, such that 
$K(A^*,B^*) \leq (1+45\eps) \cdot K^*$. 
\end{lemma}

\subsection{Proof of Lemma~\ref{lem:StructureLemma}} 
\label{subsec:StructureLemmaProof}%

Our proof begins by showing how to construct the batching $B^*=(B^*_1,\ldots,B^*_m)$ and its corresponding assignment $A^*=(A^*_1,\ldots,A^*_m)$ from the optimal schedule $\pi^*$, ensuring that $B^*$ is indeed good and that $A^*$ is indeed feasible. Subsequently, we prove the desired bound on their objective, $K(A^*,B^*) \leq (1+45\eps) \cdot K^*$. 

\paragraph{Constructing $\boldsymbol{A^*}$ and $\boldsymbol{B^*}$.} Recalling that $\chi$ is the smallest integer for which $(1 + \eps)^\chi \cdot \Delta \geq \tilde{K} \geq K^*$, we begin by setting $\beta = 2 \chi + 1 = 2 |\mathcal{B}|-1$, thereby ensuring that $B^*$ satisfies property 1 of good batchings. Next, for any day $i \in [m]$ and batch index~$b \in [\beta]$, we define the batch $(i,b)$, together with its set of assigned jobs $(A_i^*)^{-1}(b)$, depending on whether~$b$ is odd or even.
\begin{itemize}
    \item {\em Case 1: $b=2\hat{\chi}+1$ for some $\hat{\chi} \leq \chi$}. When $\hat{\chi}=0$, the set $(A_i^*)^{-1}(1)$ is defined as the set of jobs~$(i,j)$ whose completion time $C_{i,j}(\pi^*)$ is strictly smaller than $\Delta$. For $\hat{\chi} \geq 1$, the set~$(A_i^*)^{-1}(b)$ is defined as the set of jobs $(i,j)$ with a completion time of 
    \begin{equation}
    \label{case1eq}
      C_{i,j}(\pi^*) ~~\in~~ \big((1+\epsilon)^{\hat{\chi}-1} \cdot \Delta,  (1+\epsilon)^{\hat{\chi}} \cdot \Delta\big).
    \end{equation}
    In this way, the set $(A_i^*)^{-1}(1)$ consists of all jobs whose completion time is smaller than $\Delta$, which is the smallest value in ${\cal B}$. Then, $(A_i^*)^{-1}(3)$ consists of all jobs whose completion time is strictly between the smallest and second-smallest values in ${\cal B}$, so on and so forth. We set the capacity of batch $(i,b)$ to $B^*_{i,b} = \lceil \sum_{j \in (A_i^*)^{-1}(b)} p_{i,j} \rceil^{({\cal B})}$, where $\lceil \cdot \rceil^{({\cal B})}$ is an operator that rounds its argument up to the nearest number in ${\cal B}$.

    \item {\em Case 2: $b=2\hat{\chi}$ for some $\hat{\chi} \leq \chi$}. Here, the set of jobs $(A_i^*)^{-1}(b)$ assigned to batch $(i,b)$  will consist of at most one job, corresponding to the unique job $(i,j)$ for which 
    $$
    C_{i,j}(\pi^*) - p_{i,j} ~~\leq~~ (1+\eps)^{\hat{\chi}-1} \cdot \Delta ~~\leq~~ C_{i,j}(\pi^*) ~~<~~ (1+\eps)^{\hat{\chi}} \cdot \Delta.
    $$
    If such a job does not exist, we set $(A_i^*)^{-1}(b) = \emptyset$. Thus, the set $(A_i^*)^{-1}(2)$ consists of at most one job, whose processing interval contains $\Delta$, the smallest value in ${\cal B}$. Then, $(A_i^*)^{-1}(4)$ consists of at most one job, whose processing interval contains $(1+\eps) \cdot \Delta$, the second-smallest value in $\mathcal{B}$, so on and so forth. As in case~1, the capacity of this batch is set to $B^*_{i,b} = \lceil \sum_{j \in (A_i^*)^{-1}(b)} p_{i,j} \rceil^{({\cal B})}$.
\end{itemize}

Clearly, the last job scheduled by $\pi^*$ on day~$i$ will always be assigned to some batch, since by definition of~$\chi$ and~$\tilde{K}$, we have $(1+\eps)^\chi \cdot \Delta  \geq  \tilde{K}$ and $\tilde{K } \geq K^* = \max_{j \in [n]} C_{i,j}(\pi^*)$. Moreover, since each batch~$(i,b)$ has a capacity of $\lceil \sum_{j \in (A_i^*)^{-1}(b)} p_{i,j} \rceil^{({\cal B})}$, we are guaranteed that~$B^*$ only has ${\cal B}$-valued capacities, and that~$A^*$ is feasible with respect to this batching. It follows that $B^*$ satisfies both properties of good batchings, and that it admits a feasible assignment, $A^*$.

\paragraph{Bounding $\boldsymbol{K(A^*,B^*)}$.} To upper-bound $K(A^*,B^*)$ in terms of $K^*$, we will argue that for each client~$j \in [n]$ and day~$i \in [m]$, the end ${\cal E}^B_{i,A^*_i(j)}$ of the batch to which job~$(i,j)$ is assigned satisfies
\begin{equation} 
\label{eqn:StructureLemmaProof}%
{\cal E}^B_{i,A^*_i(j)} ~~\leq~~ (1 + \eps) \cdot C_{i,j}(\pi^*) + \frac{44\eps}{m} \cdot K^*.
\end{equation}
By summing this inequality across all days, it follows that 
$$
\sum_{i\in [m]} {\cal E}^B_{i,A^*_i(j)} ~~\leq~~ (1+\eps) \cdot \sum_{i \in [m]} C_{i,j}(\pi^*) + 44\eps K^* ~~=~~ (1+\eps) \cdot C_j(\pi^*) + 44\eps K^*,
$$
and therefore
$$
K(A^*,B^*) ~~=~~ \max_{j \in [n]} \sum_{i\in [m]} {\cal E}^B_{i,A^*_i(j)} ~~\leq~~  (1+\eps) \cdot \max_{j \in [n]}  C_j(\pi^*) + 44\eps K^* ~~=~~ (1+45\eps) K^*.
$$
We prove inequality~\eqref{eqn:StructureLemmaProof} by considering two cases, depending on whether job $(i,j)$ is assigned to an odd-indexed batch or to an even-indexed batch.

\paragraph{Case 1: $\boldsymbol{A^*_i(j)=b=2\hat{\chi}+1}$ for some $\boldsymbol{\hat{\chi} \leq \chi}$.} In this case, we have
\begin{eqnarray*}
{\cal E}^B_{i,A^*_i(j)} &= & \sum_{\hat{b} \leq b} B^*_{i,\hat{b}} \\
&=& \sum_{\hat{b} \leq b} \left\lceil \sum_{k \in (A^*_i)^{-1}(\hat{b})} p_{i,k} \right\rceil^{({\cal B})}\\
&\leq & \sum_{\hat{b} \leq b}\Big(  (1 + \eps) \cdot\sum_{k \in (A^*_i)^{-1}(\hat{b})}  p_{i,k} + \Delta \Big)\\
& = & (1 + \eps) \cdot \max_{ \MyAbove{k\in[n]:}{ A^*_i(k) \leq b}} C_{i,k}(\pi^*) + b \Delta,
\end{eqnarray*}
where the sole inequality above  follows from the definition of $\lceil \cdot \rceil^{({\cal B})}$, and the last equality holds since $\sum_{\hat{b} \leq b} \sum_{k\in[n]:k \in (A^*_i)^{-1}(\hat{b})} p_{i,k}$ is precisely the maximum completion time of any job assigned to one of the batches $(i,1),\ldots,(i,b)$. Now, as the latter maximum is at most $(1+\eps)^{\hat{\chi}} \cdot \Delta$, and since $C_{i,j}(\pi^*) > (1+\eps)^{\hat{\chi}-1}\cdot \Delta$, due to~\eqref{case1eq}, we have 
$$
 \max_{ \MyAbove{k\in[n]:}{ A^*_i(k) \leq b}} C_{i,k}(\pi^*) ~~\leq~~ (1+\eps)^{\hat{\chi}} \cdot \Delta  ~~<~~ (1+\eps) \cdot   C_{i,j}(\pi^*).
$$
As such, to arrive at inequality~\eqref{eqn:StructureLemmaProof}, it remains to show that $b\Delta \leq \frac{44\eps}{m} \cdot  K^*$. Indeed, recalling that $b \leq \beta$, and that $\beta \leq 2|\mathcal{B}|$ as $B^*$ is a good batching, we have 
$$
b\Delta ~~\leq~~ 2|\mathcal{B}|\Delta ~~\leq~~ 2 \cdot \frac{ 11 }{ \eps^2 } \ln m \cdot \frac{\eps^3}{m^2} \cdot (1+\eps) K^* ~~\leq~~ \frac{44\eps}{m} \cdot  K^*,
$$
where the second inequality follows from~\eqref{eqn:Delta} and \eqref{eqn:GoodBatchingLength}.

\paragraph{Case 2: $\boldsymbol{A^*_i(j)=b=2\hat{\chi}}$ for some $\boldsymbol{\hat{\chi} \leq \chi}$.} In this case, we know that $(A^*_i)^{-1}(b) = \{j\}$, i.e., $(i,j)$ was the only job assigned to the batch $(i,b)$. Thus, we have $B_{i,b}^*=  \lceil p_{i,j} \rceil^{({\cal B})}
\leq (1+\eps) \cdot p_{i,j} + \Delta$, and similarly to the analysis above,
\begin{eqnarray*}
{\cal E}^B_{i,A^*_i(j)} & = & \sum_{\hat{b} < b} B^*_{i,\hat{b}}+B_{i,b}^* \\
& \leq & (1 + \eps) \cdot \max_{ \MyAbove{ k\in[n]: }{ A^*_i(k) < b}} C_{i,k}(\pi^*) + (b-1)\Delta + (1+\eps) \cdot p_{i,j}+\Delta \\ 
&= & (1+\eps) \cdot \left( \max_{ \MyAbove{ k\in[n]: }{ A^*_i(k) < b}} C_{i,k}(\pi^*) + p_{i,j} \right) + b\Delta\\
&=& (1+\eps) \cdot C_{i,j}(\pi^*) + b\Delta \\
& \leq & (1+\eps) \cdot C_{i,j}(\pi^*) + \frac{44\eps}{m} \cdot  K^* .
\end{eqnarray*}
To see why the last equality holds, recall that the set of jobs that precede $(i,j)$ in $\pi^*_i$ is exactly those assigned to batches $(i,1),\ldots,(i,b-1)$, implying that $C_{i,j}(\pi^*)$ is precisely the sum of processing times of  these jobs plus $p_{i,j}$. The last inequality follows by noting that  $b\Delta \leq \beta\Delta \leq \frac{44\eps}{m} \cdot  K^*$, as shown in case~1. 

\subsection{The dynamic programming algorithm} 
\label{subsec:PTAS_DP}%

As explained in Section~\ref{subsec:GoodBatchings}, the number of distinct good batchings is only $2^{\tilde{O}(m/\eps^2)}$. Moreover, the entire collection of such batchings can be generated in $2^{\tilde{O}(m/\eps^2)}$ time. Among these batchings, at least one exhibits the structural properties guaranteed by Lemma~\ref{lem:StructureLemma}; we denote this batching by $B^*$. The main difficulty is that we do not know of any efficient method to compute its corresponding feasible assignment $A^*$. Nevertheless, we can efficiently compute a $\Delta$-feasible assignment~$A$ whose $K(A,B^*)$ value can be bounded in terms of $K^*$, similarly to Lemma~\ref{lem:StructureLemma}. Specifically, in the remainder of this section, we establish the next result via a dynamic programming approach.

\begin{lemma}
\label{lem:DPAlgorithm}
Given $B^*$ as input, there is an $O(n^{\tilde{O}(m / \eps^2 ) })$-time algorithm for computing a $\Delta$-feasible assignment $A$  satisfying $K(A,B^*) \leq (1+45\eps) \cdot K^*$. 
\end{lemma}

At a high level, given the good batching $B^*$, our dynamic program sequentially processes the set of clients in an arbitrary order, say from client~$1$ to client~$n$. To represent how the jobs of a single client are assigned across all $m$ days, we introduce the notion of a configuration, $c = (c_1,\ldots,c_m) \in [\beta]^m$, which is simply an $m$-dimensional vector of batch indices, capturing the concurrent decision of assigning job $(1,j)$ to batch $(1,c_1)$, job $(2,j)$ to batch $(2,c_2)$, and so on. Rather than allowing jobs to be assigned according to any possible configuration, we restrict attention to those leading to an objective value that does not deviate much above $\tilde{K}$. Specifically, for any configuration $c$, letting $K(c) = \sum_{i \in [m]} {\cal E}^B_{i,c_i}$ be the sum of batch ends implied by $c$, we say that this configuration is valid when $K(c) \leq (1+45\eps)\cdot \tilde{K}$. 

\paragraph{States and value function.} When processing the sequence of clients $1, \ldots, n$, at the end of each iteration $j\in[n]$, rather than keeping track of the precise assignment decisions for clients $1, \ldots, j$, we will only maintain a proxy for the total processing time assigned to each batch. This proxy will be created by first rounding each processing time $p_{i,j}$ down to $p_{i,j}^{\downarrow}$, standing for the nearest integer multiple of $ \frac{\Delta}{n}= \frac{\eps^3 }{nm^2} \cdot \tilde{K}$. As a result,
\begin{equation}
\label{procrelation}
p_{i,j}^{\downarrow} ~~\leq~~    p_{i,j} ~~\leq~~ p_{i,j}^{\downarrow}+\frac{\Delta}{n}.
\end{equation}

Given a partial job-to-batch  assignment, suppose that $\vec{P} = ( P_{i,b} )_{i \in [m], b \in [\beta]}$ is an $m\beta$-dimensional vector, where each coordinate $P_{i,b}$ represents the total rounded-down processing time of the jobs assigned to batch  $(i,b)$. Using this convention, at each iteration $j\in[n]$, our dynamic program will operate by evaluating the function $F(j,\vec{P})$, indicating whether there exists a sequence of valid configurations $c^{(1)}, \ldots,c^{(j)}$ for the prefix of jobs $1, \ldots, j$, such that $\sum_{k \leq j: c^{(k)}_i = b}  p_{i,k}^{\downarrow} = P_{i,b}$ for every $i \in [m]$ and $b \in [\beta]$. It is easy to verify that, since processing times were rounded down to nearest integer multiples of $\frac{\eps^3 }{nm^2} \cdot \tilde{K}$, and since we only consider valid configurations, there are $O(\frac{ nm^2 }{\eps^3 })$ possible values for each entry of the vector $\vec{P}$. 

\paragraph{Recursive equations.} Next, we explain how the value function $F(\cdot,\cdot)$ can be expressed in recursive form. To this end, terminal states of our recursion will correspond to $j=0$, in which case  $F(0,\vec{P})=\mathrm{TRUE}$ if and only if $\vec{P}=0$. For the general case of $j \geq 1$, given a vector $\vec{P}$ and a valid configuration $c$, let $\vec{P} \myominus (j,c)$ be the vector defined by 
$$
(\vec{P} \myominus (j,c))_{i,b} ~~=~~ P_{i,b} - \mathbbm{1} \{c_i=b\} \cdot p_{i,j}^\downarrow \ . $$
In other words, $\vec{P} \myominus c$ is simply the residual processing times vector incurred due to clients $1, \ldots, j-1$, when client $j$ is assigned in configuration $c$. This definition leads us to an immediate recursive formula, stating that $F(j,\vec{P}) = \mathrm{TRUE}$ if and only if there exists a valid configuration $c$ for which $F(j-1,\vec{P} \myominus (j,c))= \mathrm{TRUE}$. The resulting dynamic program proceeds by computing $F(j,\vec{P})$, for every client $j\in [n]$ and every possible vector $\vec{P}$.

Due to Lemma~\ref{lem:StructureLemma}, we know that with respect to $B^*$, there exists a feasible assignment $A^*=(A_1^*,\ldots,A_m^*)$ satisfying $K(A^*,B^*) \leq (1+45\eps) \cdot K^*$. Now, suppose that for each client $j \in [n]$, we define $c^{*(j)} = (A^*_1(j), \ldots, A^*_m(j))$. Then, $c^{*(1)},\ldots,c^{*(n)}$ is a sequence of valid configurations for jobs $1,\ldots,n$ that implies a vector $\vec{P}^*$ with $P_{i,b}^* \leq B^*_{i,b}$ in each coordinate. For this vector, we clearly have $F(n,\vec{P}^*)=\mathrm{TRUE}$.

\paragraph{Translating a TRUE solution into a $\boldsymbol{\Delta}$-feasible assignment.}
Given such a vector, $\vec{P}$,  we it can easily be translated in $O(nm)$ time into a $\Delta$-feasible assignment $A$ with respect to $B^*$ such that $K(A,B^*)\leq(1+45\eps) \cdot \tilde{K}$. For this purpose, by standard backtracking techniques, we first recover the sequence of valid configurations $c^{(1)},\ldots,c^{(n)}$ leading to $F(n,\vec{P})=\mathrm{TRUE}$. Given these configurations, we construct a corresponding assignment $A$ such that $A_i(j)=c_i^{(j)}$ for each client $j \in [n]$ and day $i \in [m]$. With this definition, 
\begin{eqnarray} 
&& K(A,B^*) ~~=~~ \max_{j \in [n]} \sum_{i\in [m]} {\cal E}^B_{i,A_i(j)} ~~=~~ \max_{j \in [n]} \sum_{i\in [m]} {\cal E}^B_{i,c_i^{(j)}} \nonumber \\
&& \qquad \quad \qquad \qquad \qquad \qquad \qquad =~~ K( c^{(j)} ) ~~\leq~~ (1+45\eps)\cdot \tilde{K} \ ,    \label{eqn:UB_KABstar}
\end{eqnarray}
where the last inequality holds since $c^{(j)}$ is valid. That said, even though $P_{i,b} \leq B^*_{i,b}$ for all $i \in [m]$ and $b \in [\beta]$, the assignment $A$ may be infeasible, since the entries of $\vec{P}$ are expressed in terms of the rounded-down processing times $\{p_{i,j}^\downarrow\}_{i \in [m], j \in [n]}$. However, due to relation~\eqref{procrelation} between $p_{i,j}^\downarrow$ and $p_{i,j}$, when reverting back to the original processing times, each  batch~$(i,b)$ has a total processing time of
$$ 
\sum_{ \MyAbove{j\in[n]:}{A_i(j) = b}} p_{i,j} ~~\leq~~ \sum_{\MyAbove{j\in[n]:}{A_i(j) = b}} \Big(  p_{i,j}^{\downarrow} + \frac{\Delta}{n} \Big) ~~\leq~~ P_{i,b} + \Delta ~~\leq~~ B^*_{i,b} + \Delta, 
$$
meaning that our assignment is guaranteed to be $\Delta$-feasible.

\paragraph{Running time.} To complete the proof of Lemma~\ref{lem:DPAlgorithm}, let us analyze the running time of this procedure. First, since $B^*$ is a good batching, its length is $\beta =O(\frac{\log m}{\eps^2})$. Next, there are only $O(\frac{nm^2}{\eps^3})$ possible values for each entry $P_{i,b}$, leading to $O((\frac{nm^2}{\eps^3})^{m\beta}) = O( n^{\tilde{O}(m/\eps^2)} )$ distinct vectors $\vec{P}$. In addition, the total number of valid configurations is at most $\beta^m = 2^{ \tilde{O}( m \log \frac{1}{\eps} ) }$, meaning that our dynamic program requires  $O( n^{\tilde{O}(m/\eps^2)} )$ time. 

\paragraph{Conclusion.} As previously shown in Lemma~\ref{lem:AssignmentToSchedule}, since the assignment $A$ is $\Delta$-feasible, it can be translated in $O(mn)$ time to a schedule $\pi$ whose maximal sum of completion times is $K(\pi) \leq K(A,B^*) + m \beta\Delta$. To bound this expression in terms of $K^*$, we observe that
\begin{eqnarray*}
K(\pi) &\leq & K(A,B^*) + m \beta\Delta \\
& \leq & (1+45\eps) \cdot \tilde{K} +  22\eps \tilde{K} \\ 
& \leq &(1+67\eps)(1+\eps) \cdot K^* \\
& \leq & (1+135\eps) \cdot K^*.
\end{eqnarray*}
Here, the second inequality holds since $K(A,B^*) \leq (1+45\eps) \cdot \tilde{K}$, by inequality~\eqref{eqn:UB_KABstar}. In addition, $\beta \leq 2\cdot |{\cal B}| \leq \frac{ 22 \ln m }{ \eps^2 }$ according to~\eqref{eqn:GoodBatchingLength}, and $\Delta = \frac{\eps^3}{m^2} \cdot \tilde{K}$ by definition~\eqref{eqn:Delta}, implying that $m \beta\Delta \leq 22\eps \tilde{K}$. The third inequality holds since $\tilde{K} \leq (1+\eps)K^*$.

To bound the overall running time of our algorithm, we observe that it is dominated by the dynamic programming procedure of  Lemma~\ref{lem:DPAlgorithm}, which is employed for every possible good batching. As mentioned in Section~\ref{subsec:GoodBatchings}, there are only~$2^{\tilde{O}(m/ \eps^2)}$ good batchings to consider, all generated in~$2^{\tilde{O}(m/ \eps^2)}$ time. Given that our dynamic program admits an $O(n^{\tilde{O}(m/\eps^2)})$-time implementation, we arrive at a total running time of $O(n^{\tilde{O}(m/\eps^2)})$. 
\section{The Day-Invariant Model: \texorpdfstring{$\boldsymbol{(\frac{1+\sqrt{2}}{2}+\epsilon)}$}{}-Approximation} \label{sec:identical_better}

In this section, we consider the day-invariant model, where each client $j\in[n]$ has a uniform  processing time of $p_{i,j}=p_j$ across all $m$ days. We show that this unique structure can be exploited to improve on our $2$-approximation for the day-dependent model. As formally stated in Theorem~\ref{thm:BetterApprox}, we prove that for any $\epsilon > 0$, day-invariant instances can be approximated in $O(n^{\tilde{O}(1/\eps^3)})$ time within factor $\frac{1+\sqrt{2}}{2}+\epsilon$ of optimal. Toward this objective, in Sections~\ref{subsec:EnhancedLowerBound} and~\ref{subsec:TwoDayInversion}, our analysis begins by deriving a new lower bound on the optimal objective value, which motivates us to propose the two-day inversion algorithm. The performance of this approach will be analyzed in Section~\ref{subsec:iden_const_analysis} via a direct comparison to our new bound.

\subsection{Lower-bounding method}
\label{subsec:EnhancedLowerBound}%

Throughout this section, $P=\sum_{j\in[n]} p_j$ will denote the total processing time of all jobs on a single day, whereas  $p_{\max}=\max_{j\in[n]} p_j$ will stand for the maximal processing time of any job. With this notation, we prove that day-invariant processing times allow us to lower-bound the optimum value $K^*$ of any instance in terms of $m$, $P$, and $p_{\max}$ along the lines of the following claim. 

\begin{lemma}
\label{lem:EnhancedLowerBound}
$
K^* \ge \frac{m}{2} \cdot (P+\frac{p_{\max}^2}{P}).
$
\end{lemma}

\paragraph{The effect of client-splitting.} To derive this bound, we observe that by splitting any client into a subset of clients with an identical total processing time, the optimum value can only decrease. For this purpose, consider an arbitrary day-invariant instance $I$, and suppose that client~$j$ has a job processing time of $p_j \geq 2$. Since $p_j$ is assumed to be an integer, we construct a new instance~$\tilde{I}$, in which  client $j$ is replaced by two clients, $j_1$ and $j_2$, with  $p_{j_1} = 1$ and $p_{j_2} = p_j -1$. It is easy to verify that the optimal solution value of $I$ is lower-bounded by that of $\tilde{I}$. Indeed, without increasing its objective value, any schedule $\pi$ for $I$ can be converted into a schedule $\tilde{\pi}$ for~$\tilde{I}$ by placing $j_1$'s job in the same position as $j$'s job with respect to the schedule $\pi_i$ of each day $i$. Then, $j_2$'s job is inserted immediately after, shifting the jobs of all subsequent clients one position forward.

\paragraph{Splitting to unit processing times.} With this observation, given an instance $I$ of the day-invariant model, let $\ell \in [n]$ be a client whose job attains the maximal processing time across all clients, i.e., $p_{\ell} = p_{\max}$. We proceed by reducing $I$ to an instance~$\tilde{I}$, where every client $j \neq \ell$ is split into $p_j$ clients with unit processing times. In contrast, the jobs associated with client $\ell$ remain unchanged, meaning that the resulting instance has $P-p_{\max}+1$ clients. As explained above, the optimum $K^*$ of~$I$ is lower-bounded by the optimum value $\tilde{K}$ of~$\tilde{I}$, and we proceed to derive Lemma~\ref{lem:EnhancedLowerBound} by arguing that $\tilde{K} \ge \frac{m}{2} \cdot (P+\frac{p_{\max}^2}{P}).$

\paragraph{Lower-bounding $\boldsymbol{\tilde{K}}$.} Next, let $\tilde{\pi}$ be an optimal schedule with respect to $\tilde{I}$, and for the unsplit client $\ell$, let $x_i \in [p_{\max},P]$ be the completion time of job $(i,\ell)$, for each day $i\in[m]$. Our first lower bound on $\tilde{K}$ is straightforward: Since $C_{\ell}( \tilde{\pi} ) = \sum_{i\in[m]} x_i$ is the total completion time of client~$\ell$, we must have $\tilde{K} \geq \sum_{i\in[m]} x_i$. The next claim establishes the complementary non-trivial bound, by accounting for the remaining clients. 

\begin{lemma}
\label{lem:Restricted}%
$\tilde{K} \geq \sum_{i \in [m]}\frac{P^2-2p_{\max}x_i+p_{\max}^2}{2(P-p_{\max})}$.
\end{lemma}
\begin{proof}
Focusing on a single day $i \in [m]$, we recall that by definition of $x_i$, job $(i,\ell)$ is necessarily scheduled along the time interval $(x_i - p_{\max},x_i]$. Therefore, all other jobs on this day are scheduled either during the ``lower'' interval $(0,x_i - p_{\max}]$ or during the ``upper'' interval $(x_i,P]$. Since these jobs have unit processing times,  $x_i - p_{\max}$ jobs are scheduled in the lower interval. Consequently, regardless of processing order, their total completion time is $\frac{1}{2} \cdot (x_i - p_{\max}) (x_i - p_{\max} +1)$. Similarly, $P-x_i$ jobs are scheduled in the upper interval, with a total completion time of $\frac{1}{2} \cdot (P-x_i)(P+x_i+1)$.

Thus, over all $m$ days, the total  completion time of all jobs other than $(i,\ell)$ is 
\begin{eqnarray*}
&& \frac{ 1 }{ 2 } \cdot \sum_{i \in [m]} \left( (x_i - p_{\max}) (x_i - p_{\max} +1)+ (P-x_i)(P+x_i+1) \right) \\
&& \qquad >~~ \frac{ 1 }{ 2 } \cdot \sum_{i \in [m]}  \left( (x_i-p_{\max})^2+(P-x_i)(P+x_i) \right)\\
&& \qquad =~~ \frac{ 1 }{ 2 } \cdot \sum_{i \in [m]} \left( P^2-2p_{\max}x_i+p_{\max}^2 \right) .       
\end{eqnarray*}
The desired claim thus follows by  observing that the maximal total completion time $\tilde{K}$ is lower-bounded by the average total completion time over all $P - p_{\max}$ jobs, and therefore, $\tilde{K} \geq \sum_{i \in [m]}\frac{P^2-2p_{\max}x_i+p_{\max}^2}{2(P-p_{\max})}$. 
\end{proof}

Now, letting $x=\frac{1}{m} \cdot \sum_{i\in[m]}x_i$, by combining our first lower bound of $\tilde{K} \geq \sum_{i\in[m]} x_i$ with that of Lemma~\ref{lem:Restricted}, we get 
\begin{eqnarray*}
\frac{\tilde{K}}{m} &\geq&  \max \left\{x,\frac{P^2-2p_{\max}x+p_{\max}^2}{2(P-p_{\max})} \right\} \\
& \geq & \min_{\hat{x} \in[p_{\max},P]}  \max\left\{\hat{x},\frac{P^2-2p_{\max}\hat{x}+p_{\max}^2}{2(P-p_{\max})} \right\} \\
& = & \frac{1}{2} \cdot \Big(P+\frac{p_{\max}^2}{P}\Big).   
\end{eqnarray*}

\subsection{The two-day inversion algorithm}
\label{subsec:TwoDayInversion}%

Given $\eps > 0$, our algorithm proceeds by considering two cases, depending on how the number of days $m$ is related to the accuracy parameter $\eps$. The first case corresponds to having very few days, specifically meaning that $m < \frac{ 1 }{ \eps }$. Here, we simply employ the approximation scheme presented in Section~\ref{sec:PTAS}, ending up with a $(1+\epsilon)$-approximate schedule  in $O(n^{\tilde{O}(m/\eps^2)})=O(n^{\tilde{O}(1/\eps^3)})$ time. Moving forward, we consider the non-trivial many-days case, where $m \geq \frac{ 1 }{ \eps }$

To handle this scenario, the two-day inversion algorithm fixes an arbitrary order, say $1, \ldots, n$, and alternates between this order and its inverse on odd and even days. Formally, our schedule $\pi=(\pi_1,\ldots,\pi_m)$ sets $\pi_i(j)=j$ for all clients $j \in [n]$ when day $i$ is odd, and $\pi_i(j)=n-j+1$ when day $i$ is even. As a result, every client $j \in [n]$ incurs a completion time of $\sum_{k\leq j}p_{k}$ on any odd day and $\sum_{k\ge j}p_{k}$ on any even day. By summing these completion times on any pair of successive days, we get $P+p_j$, implying that the resulting schedule has a maximum total completion time of
\begin{equation}
\label{eqn:HeuristicSolution}
K(\pi) ~~\leq~~
\left \lfloor \frac{m}{2} \right  \rfloor \cdot (P+p_{\max}) + P. 
\end{equation}

\subsection{Analysis} \label{subsec:iden_const_analysis}

We claim that the schedule $\pi$ produced by the two-day inversion algorithm provides an approximation guarantee of $\frac{1+\sqrt{2}}{2} + O(\epsilon)$, thereby completing the proof of Theorem~\ref{thm:BetterApprox}. To this end, according to inequality~\eqref{eqn:HeuristicSolution}, we know that
\[ K(\pi) ~~\leq~~ \frac{m}{2} \cdot (P+p_{\max}) + P ~~\leq~~ \frac{m}{2} \cdot (P+p_{\max}) + 2\epsilon K^* , \]
where the last inequality holds since $K^* \geq \frac{m}{2} \cdot (P+\frac{p_{\max}^2}{P})
$ by Lemma~\ref{lem:EnhancedLowerBound}, implying in particular that $P \leq \frac{ 2K^* }{ m } \leq 2\eps K^*$, as $m \geq \frac{ 1 }{ \eps }$. 

Therefore, we can bound the maximum possible approximation ratio $\frac{ K(\pi) }{ K^* }$ by finding $P$ and $p_{\max}$ values that optimize  the following mathematical program:
\begin{alignat}{3}
& \text{max} &\quad & \frac{m\cdot (P+p_{\max})}{2K^*} \nonumber   \\
&\text{s.t.}
&\quad & K^* \ge \frac{m}{2} \cdot\Big(P+\frac{p_{\max}^2}{P} \Big) \label{eqn:2day_const_1} \\
&&\quad & K^* \geq m  p_{\max} \label{eqn:2day_const_2}\\
&&\quad & P\ge p_{\max} \ge 0 \label{eqn:2day_const_3} 
\end{alignat}
In this formulation, constraints~\eqref{eqn:2day_const_1} and~\eqref{eqn:2day_const_2} follow from Lemma~\ref{lem:EnhancedLowerBound}, stating that $K^* \ge \frac{m}{2} \cdot (P+\frac{p_{\max}^2}{P})
$, and from the observation that $m  p_{\max}$ is a trivial lower bound on $K^*$. Constraint~\eqref{eqn:2day_const_3} directly follows from how $P$ and $p_{\max}$ are related. Now, by exchanging our decision variables with $x=\frac{mP}{K^*}$ and $y=\frac{mp_{\max}}{K^*}$, we obtain a simplified formulation, given by
\begin{alignat*}{3}
& \text{max} &\quad & \frac{x+y}{2} \\
&\text{s.t.}
&\quad & x+\frac{y^2}{x} \leq 2 \\
&&\quad & y \leq 1\\
&&\quad & x\ge y \ge 0 
\end{alignat*}
It is not difficult to verify that the optimal solution to this formulation is $(x^*,y^*)=(1+ \frac{ 1 }{ \sqrt2 }, \frac{ 1 }{ \sqrt2})$, meaning that $\frac{ K(\pi) }{ K^* } \leq \frac{ x^* + y^* }{ 2 } + 2\epsilon=\frac{1+\sqrt{2}}{2} + 2\epsilon$.
\section{The Day-Invariant Model: Quasi-PTAS}
\label{sec:QPTAS}%

In this section, we present a quasi-polynomial-time approximation scheme for the day-invariant model. As stated in Theorem~\ref{thm:QPTAS}, we will show that, for any $\eps > 0$, this model is approximable within factor $1+\eps$ of optimal. Our algorithm admits an $O(n^{\tilde{O}( \frac{ \log \log n }{ \eps^9 })})$-time implementation, sitting at the very low end of quasi-polynomial time.

\paragraph{Outline.} The starting point of our approach is established in Section~\ref{subsec:reduction_days}, where we propose a general reduction from $m$ days to $O(\frac{\log n}{\eps^3})$ days, losing a factor of only $1 + O(\eps)$ in optimality. Coupled with Theorem~\ref{thm:PTAS}, this result already yields an $O(n^{\tilde{O}(\frac{\log n}{\eps^5})})$-time approximation scheme for the day-invariant model. To obtain an improved running time of $O(n^{\tilde{O}( \frac{ \log \log n }{ \eps^9 })})$, Section~\ref{subsec:SpecializedGoodBatchings} introduces a refinement of good batchings, tailored to the simpler structure of day-invariant processing times. Then, in Sections~\ref{subsec:LPforComputingA}-\ref{subsec:bound_stretch}, we employ randomized rounding with respect to a new configuration LP, for the purpose of computing an approximate job-to-batch assignment, bypassing the dynamic programming bottleneck that is inherent to Theorem~\ref{thm:PTAS}. 

\subsection{Approximate reduction to \texorpdfstring{$\boldsymbol{O(\frac{ \log n }{ \eps^3})}$}{} days} \label{subsec:reduction_days}

In what follows, we show how to approximately reduce any $m$-day instance of the day-invariant model to an $O(\frac{ \log n }{ \eps^3} )$-day instance, losing a factor of $1 + O(\eps)$ in optimality. This result will be our gateway to the design of a slightly-superpolynomial approximation scheme and to a complete proof of Theorem~\ref{thm:QPTAS} later on. 

For the remainder of this section, let $K^*$ be the optimum value of our given instance. Without loss of generality, we assume that $m \geq \frac{\log n}{\eps^3}$; otherwise, no modifications are required. Letting $D= \lceil \frac{ 6\ln (2n) }{ \eps^2} \rceil$, the type of reduction we propose can be formalized as follows. 

\begin{lemma} 
\label{lem:SmallDaysReduction}%
Let $\pi_D$ be an $\alpha$-approximate schedule for $D$ days, for some $\alpha \ge 1$. Then, $\pi_D$ can be translated in $O(nm)$ time to a schedule $\pi$ for the entire $m$-day instance such that
$$
K(\pi) ~~\leq~~ (1 + 26\eps) \cdot \alpha K^*.
$$
\end{lemma}

Toward proving this result, suppose that $K^*_D$ stands for the optimum value of a $D$-day instance. In Lemma~\ref{lem:opt_d_vs_m}, we  show that the optimum-to-day ratio $\frac{ K^*_D }{D}$ does not deviate much above $\frac{ K^* }{ m }$. This relation will allow for a relatively simple reduction, where we essentially replicate the schedule $\pi_D$ across the entire period of $m$ days. Our proof works by employing the probabilistic method~\cite{AlonS2016}, where the next Chernoff-Hoeffding bound will be useful; see Dubhashi and Panconesi~\cite[Thm.~1.1 and Ex.~1.1]{DubhashiP09} for its proof.

\begin{lemma}
\label{lem:Chernoff-Hoeffding}%
Let $Z_1, \ldots, Z_n$ be independent $[0,1]$-bounded random variables. Then, for every $0 < \eps < 1$ and $\bar{\mu} \geq \expar{ \sum_{i \in [n]} Z_i }$,
$$
\pr{ \sum_{i \in [n]} Z_i \ge (1 + \eps) \cdot \bar{\mu}  } ~~\leq~~ \exp \left( - \frac{ \eps^2 \bar{\mu} }{ 3 } \right).
$$
\end{lemma}

\begin{lemma} 
\label{lem:opt_d_vs_m}%
$\frac{K^*_D}{ {D} } \leq (1 + \eps) \cdot \frac{K^*}{ m }$.
\end{lemma}
\begin{proof}
Let $\pi^*$ be an optimal $m$-day schedule, with $\max_{j \in [n]} C_j(\pi^*) = K^*$. Suppose we create a random $D$-day schedule $\pi_D$ by independently sampling $D$ days from $\pi^*$, each chosen uniformly at random with replacement. We claim that, with positive probability, $C_j(\pi_D) \leq (1 + \eps) \cdot D \cdot \frac{K^*}{m}$ holds simultaneously for all clients $j \in [n]$, implying that $K^*_D \leq (1 + \eps) \cdot D \cdot \frac{K^*}{m}$, as desired. For this purpose, by employing the union bound, it suffices to show that $\prpar{ C_j(\pi_D) \geq (1 + \eps) \cdot D \cdot \frac{ K^* }{ m } } \leq \frac{ 1 }{2n}$ for every client $j \in [n]$.  

To derive the latter bound, let $\delta_1, \ldots, \delta_D \in [m]$ be the random collection of sampled days, which are mutually independent. Since each day $\delta_{{d}}$ is sampled uniformly at random, the completion time $C_{\delta_{{d}}, j}(\pi^*)$ of job $j$ on this day, which is obviously $[0,P]$-bounded, has an expected value of
$$ 
\ex{C_{\delta_{{d}}, j}(\pi^*)} ~~=~~ \frac{ 1 }{ m } \cdot \sum_{i \in [m]} C_{i,j}(\pi^*) ~~=~~ \frac{ 1 }{ m } \cdot C_j(\pi^*) ~~\leq~~ \frac{K^* }{ m } \ . $$
As a result, since the (random) total completion time incurred by client $j$ is $C_j(\pi_D)=\sum_{{d}\in [D]}C_{\delta_{{d}}, j}(\pi^*)$, we have
\begin{eqnarray*}
\pr{ C_j(\pi_D) \geq (1 + \eps) \cdot D \cdot \frac{ K^* }{ m } } & = & \pr{ \sum_{{{d}\in [D]}} \frac{ C_{\delta_{{d}}, j}(\pi^*) }{ P } \geq (1 + \eps) \cdot D \cdot \frac{K^*}{mP} } \\
& \leq & \exp \left( - \frac{ \eps^2  }{ 3 } \cdot \frac{ DK^* }{ mP } \right) \\
& \leq &  \frac{1}{2n} .
\end{eqnarray*}
Here, the first inequality follows from the Chernoff-Hoeffding bound of Lemma~\ref{lem:Chernoff-Hoeffding}, instantiated with $Z_{\delta_{{d}}} = \frac{C_{\delta_{{d}},j}(\pi^*)}{P}$ for each ${d} \in [D]$, and $\bar{\mu} = \frac{DK^*}{mP}$. The second inequality is obtained  by recalling that $D = \lceil \frac{ 6\ln (2n) }{ \eps^2}  \rceil$ and that $K^* \geq \frac{mP}{2}$, as implied by Lemma~\ref{lem:EnhancedLowerBound}.  
\end{proof}

\paragraph{Concluding the proof of Lemma~\ref{lem:SmallDaysReduction}.} Given this relation between $K^*_D$ and $K^*$, our reduction operates as follows. Let $\pi_D$ be an $\alpha$-approximate schedule for $D$ days. Then, to construct an $m$-day schedule $\pi$, we replicate this schedule $\lfloor \frac{m}{D} \rfloor$ times, filling the remaining $m - \lfloor \frac{m}{D} \rfloor \cdot D$ days with  arbitrary schedules. To complete the proof, we argue that $C_j(\pi) \leq (1 + 26\eps) \cdot \alpha K^*$ for any client $j \in [n]$. Indeed,
\begin{eqnarray}
C_j(\pi) & \leq & \left\lfloor \frac{m}{D} \right\rfloor \cdot C_j(\pi_D) + \left( m - \left\lfloor \frac{m}{D} \right\rfloor \cdot D \right) \cdot P \nonumber \\
& \leq & \frac{ m }{D} \cdot \alpha \cdot K^*_D + D P \label{eqn:reduc_end_1} \\
& \leq & (1+\eps) \cdot \alpha K^* + \frac{D}{ m } \cdot mP \label{eqn:reduc_end_2} \\
& \leq & (1+\eps) \cdot \alpha K^* + \frac{ \lceil \frac{ 6\ln (2n) }{ \eps^2} \rceil }{ \frac{ \ln n }{ \eps^3}  } \cdot 2K^* \label{eqn:reduc_end_3} \\
& \leq & (1+26\eps) \cdot \alpha K^*. \nonumber
\end{eqnarray}
Here, we arrive at inequality~\eqref{eqn:reduc_end_1} by noting that $\pi_D$ is an $\alpha$-approximate schedule for $d$ days. Inequality~\eqref{eqn:reduc_end_2} holds since $\frac{K^*_D}{ {D} } \leq (1 + \eps) \cdot \frac{K^*}{ m }$, by Lemma~\ref{lem:opt_d_vs_m}. Finally, inequality~\eqref{eqn:reduc_end_3} is obtained by recalling that $m \geq \frac{ \ln n }{ \eps^3}$, $D = \lceil \frac{ 6\ln (2n) }{ \eps^2} \rceil$, and $K^* \geq \frac{ mP }{ 2 }$ by Lemma~\ref{lem:EnhancedLowerBound}.

\paragraph{Intermediate summary.} At the moment, let us observe that  Lemma~\ref{lem:SmallDaysReduction} leads by itself to an $O(n^{\tilde{O}(\frac{\log n}{\eps^5})})$-time approximation scheme for the day-invariant model. Indeed, since we are now dealing with an $O(\frac{ \log n }{ \eps^3} )$-day instance, one can naively employ our approximation scheme for the day-dependent model, whose specifics are summarized in Theorem~\ref{thm:PTAS}. By instantiating this result with $m = O(\frac{ \log n }{ \eps^3} )$ days, we indeed end up with a $(1+\eps)$-approximation in $O(n^{\tilde{O}(\frac{\log n}{\eps^5})})$ time.

\subsection{Refined good batchings} 
\label{subsec:SpecializedGoodBatchings}

Moving forward, our objective is to improve the above-mentioned logarithmic exponent, which seems inherent to the dynamic programming approach in play. To this end, we will refine the notion of good batchings to day-invariant processing times, and bypass the dynamic programming bottleneck by replacing it with randomized rounding of a configuration LP.

\paragraph{Definitions and notation.} As discussed in Section~\ref{sec:PTAS}, a batching $B=(B_1,\ldots,B_m)$ is simply a sequence of $m$ daily batchings, where each $B_i=(B_{i,1},\ldots,B_{i,\beta})$ consists of $\beta$ batches. In turn, an assignment $A=(A_1,\ldots,A_m)$ is a sequence of functions $A_i:[n]\to[\beta]$,  specifying how the underlying jobs are assigned to batches on each day~$i$. The objective value $K(A,B)$ of a batching $B$, along with a corresponding assignment $A$, was defined as $K(A,B) = \max_{j \in [n]} \sum_{i \in [m]} {\cal E}^B_{i,A_i(j)}$. Finally, given the approximate reduction described in Section~\ref{subsec:reduction_days}, we assume from this point on that our instance consists of $m\leq \frac{\ln n}{\eps^3}$ days. 

In analogy to Section~\ref{subsec:GoodBatchings}, let us begin by defining the discrete set of values
$$
{\cal B} ~~=~~ \Big\{ \eps^3 P, (1 + \eps) \cdot \eps^3 P, (1 + \eps)^2 \cdot \eps^3 P, \ldots, (1 + \eps)^\chi \cdot \eps^3 P \Big\}. 
$$
Here, $\chi$ is the smallest integer for which $(1 + \eps)^\chi \cdot \eps^3 \geq 1$, implying that 
\begin{equation}
\label{eqn:RefinedGoodBatchingLength}%
|{\cal B}| ~~=~~ \left \lceil \log_{1+\eps} \left(\frac{ 1 }{ \eps^3 } \right) \right \rceil ~~\leq~~ \frac{ 6 }{ \eps } \ln \frac{ 1 }{ \eps }.    
\end{equation}
In turn, we remind readers that a batching $B$ is called ``good'' when, for every day $i \in [m]$, its number of batches $\beta$ is at most $2|\mathcal{B}|$, and in addition, $B_{i,b} \in {\cal B}$ for every $b \in [\beta]$. As such, according to our new definition of ${\cal B}$, the overall number of good batchings is only 
\begin{equation}
\label{eqn:NumberofRefinedGoodBatchings}%
O( |{\cal B}|^{\beta \cdot m} )  ~~=~~  O( |{\cal B}|^{ O(|{\cal B}| \cdot m) } )  ~~=~~ \left( \tilde{O} \left(\frac{1}{\eps}\right) \right)^{\tilde{O} (\frac{\log n}{\eps^4})} ~~=~~ O(n^{ \tilde{O}(1/ \eps^4)}),    
\end{equation}
where the second equality holds since $\beta \leq 2|{\cal B}|=\tilde{O}(\frac{1}{\eps})$ and  $m=O(\frac{\log n}{\eps^3})$. 

\paragraph{The adapted structure lemma.} A fundamental difference between the definitions above and those appearing in the general context of Section~\ref{sec:PTAS} is that the cardinality of our current set ${\cal B}$ of possible batch capacities solely depends  on $1/\eps$. Nevertheless, we can still prove an analog of the day-dependent structure lemma (Lemma~\ref{lem:StructureLemma}), showing that even when we restrict ourselves to good batchings as defined above, there exists some good batching $B$ along with a feasible assignment $A$ for which $K(A,B)$ nearly matches the optimum value $K^*$. Since the proof of this result is similar in spirit to that of Lemma~\ref{lem:StructureLemma}, we present its details in Appendix~\ref{app:ProofRefinedStructure}.

\begin{lemma}
\label{lem:RefinedStructureLemma}%
There exists a good batching $B^*$, together with a feasible assignment $A^*$, such that $K(A^*,B^*) \leq (1+29\eps) \cdot K^*$.
\end{lemma}

\subsection{The configuration LP}
\label{subsec:LPforComputingA}%

Similarly to Section~\ref{sec:PTAS}, our algorithm begins by guessing the good batching $B^*$ whose existence is guaranteed by Lemma~\ref{lem:RefinedStructureLemma}. As previously noted, there are only $O(n^{\tilde{O}(1/\eps^4)})$ good batchings to consider (see equation~\eqref{eqn:NumberofRefinedGoodBatchings}). By enumerating over these options, we may assume that $B^*$ is available, incurring a multiplicative term of $n^{\tilde{O}(1/\eps^4)}$ in running time. The major improvement in our approach comes from devising a more efficient algorithm for computing the job-to-batch  assignment $A$ corresponding to $B^*$. Unlike Section~\ref{sec:PTAS}, we will ensure that this assignment has a multiplicative overflow in the capacity of each batch, rather than an additive one. In addition, the dynamic programming procedure behind our earlier approach will be replaced by LP-rounding ideas.

\paragraph{Configurations and client types.} To introduce our linear program, we remind readers  that a configuration is an $m$-dimensional vector $c = (c_{1},\ldots,c_m) \in [\beta]^m$, where each entry $c_{i}$ indicates the batch to which a given client’s job is assigned on day~$i$. For any such configuration, let $K(c) = \sum_{i \in [m]} {\cal E}^B_{i,c_{i}}$ be the sum of batch ends implied by~$c$. In light of Lemma~\ref{lem:RefinedStructureLemma}, we say that $c$ is valid when $K(c) \leq (1+29\eps)\cdot K^*$, and we make use of ${\cal C}$ to denote the set of all valid configurations, noting that 
\begin{equation}
\label{complexitycalc}
|{\cal C}| ~~\leq~~ \beta^{m} ~~=~~ {O}\Big(\frac{1}{\eps} \log \frac{1}{\eps}\Big)^{O(\frac{\log n}{\eps^3})} ~~=~~O(n^{\tilde{O}( 1/\eps^3)}).
\end{equation}

Next, letting $\Lambda = \frac{ \eps^6 P }{ 6  \ln(2m\beta)}$, we say that client $j \in [n]$ is large when $p_j \geq \Lambda$;
otherwise, this client is small. We use $L$ and $S$ to respectively denote the sets of large and small clients. Clearly, since $P = \sum_{j \in [n]} p_j$, we must have
\begin{equation}
\label{numberlarge}
|L| ~~\leq~~ \Big\lceil \frac{ 6  \ln(2m\beta) }{ \eps^6 } \Big\rceil ~~=~~ \tilde{O}\Big( \frac{ \log \log n }{ \eps^6 } \Big),
\end{equation}
where the last equality holds since $m = O( \frac{ \log n }{ \eps^3 } )$ and $\beta = \tilde{O}( \frac{ 1 }{ \eps } )$. Now, let $A^*$ be the feasible assignment corresponding to~$B^*$, whose existence is promised by Lemma~\ref{lem:RefinedStructureLemma}, and let $c^{*(j)} = (A^*_1(j), \ldots, A^*_{m}(j))$ be the configuration chosen by $A^*$ for each client~$j \in [n]$, which is necessarily valid. The next step of our algorithm consists of guessing the configuration $c^{*(j)}$ of every large client $j \in L$. Given inequalities~\eqref{complexitycalc} and~\eqref{numberlarge}, the number of guesses to examine is  $|{\cal C}|^{|L|}= O(n^{ \tilde{O}( \frac{ \log \log n }{ \eps^9 } ) })$. 

\paragraph{Formulating the configuration LP.} At this stage, we are ready to describe our feasibility linear program, \mylp, which will guide us in determining the configurations of small clients, given those of large clients, $\{c^{*(j)}\}_{j \in L}$, that have already been guessed. This program has a decision variable $x_{j,c}$ for every client~$j \in [n]$ and valid configuration $c \in {\cal C}$. In an integer-valued solution, $x_{j,c}$ should be interpreted as indicating whether client~$j$ is assigned according to configuration~$c$ or not. As such, \mylp consists of the following four sets of constraints:
\begin{alignat}{3}
& &\quad & \sum_{c \in {\cal C}}  x_{j,c} = 1 & \qquad \qquad & \forall \, j \in [n] \label{one} \\
& &\quad & \sum_{j \in [n]} \sum_{c\in {\cal C}: c_i = b } p_{j} x_{j,c} \leq B^*_{i,b} & \qquad \qquad & \forall \, i \in [m], \, b \in [\beta] \label{second} \\
& &\quad & x_{j,c^{*(j)}} = 1 & \qquad \qquad & \forall \, j \in L \label{three} \\
& &\quad & x_{j,c} \geq 0 & \qquad \qquad & \forall \, j \in [n], \, c \in {\cal C} \label{four}
\end{alignat}
Here, constraint~(\ref{one}) forces us to choose a single configuration for each client, while constraint~(\ref{second}) guaranties that the capacity of each batch is not exceeded. Constraint~(\ref{three}) ensures that each large client $j$ complies with our guessing procedure, i.e., it is assigned according to configuration $c^{*(j)}$. Finally, constraint~(\ref{four}) relaxes the requirement that each $x_{j,c}$ takes a binary value. 

\paragraph{Feasibility.} To verify that \mylp is indeed feasible, let $x^{A^*}$ be the solution induced by~$A^*$, where  $x^{A^*}_{j,c}=1$ if and only if $c = c^{*(j)}$, for each client $j \in [n]$ and configuration $c \in {\cal C}$. Here, it is important to note that each $c^{*(j)}$ is valid, since $K(A^*,B^*) \leq (1+29\epsilon) \cdot K^*$ by our refined structure lemma. Moreover, it is not difficult to verify  that~$x^{A^*}$ is feasible for \mylp, by using this lemma once again. Specifically, constraint~\eqref{second} is satisfied since $A^*$ is a feasible assignment for $B^*$, while the three remaining constraints are satisfied by definition. It follows that \mylp is feasible, and since it consists of $O(n^{\tilde{O}( 1/\eps^3)})$ variables and constraints, we can compute a feasible solution $x^*$ in polynomial time. 

\subsection{Rounding \texorpdfstring{$\boldsymbol{x^*}$}{} to a random schedule}
\label{subsec:ConstructingA}

\paragraph{Randomized rounding.} According to constraints~\eqref{one} and~\eqref{four}, it follows that $\sum_{c \in {\cal C}} x^*_{j,c} = 1$ for every client $j \in [n]$, and that each variable $x^*_{j,c}$ is non-negative, meaning that $\{ x^*_{j,c} \}_{c \in {\cal C}}$ can be viewed as a distribution over the collection ${\cal C}$ of valid configurations. Based on this distribution, we define a random valid configuration $R^{(j)}$ for each client $j$, independently drawn from ${\cal C}$ according to the probabilities $\{ x^*_{j,c} \}_{c \in {\cal C}}$. We then construct an assignment $A$ corresponding to the outcomes $R^{(1)},\ldots,R^{(n)}$, such that $A_i(j) = R^{(j)}_i$ for every client $j \in [n]$ and day $i \in [m]$; in other words, each job $(i,j)$ is placed by $A$ in batch $(i,R^{(j)}_i)$. 

\paragraph{Bounding $\boldsymbol{K(A,B^*)}$.} We first argue that, with probability $1$, the random assignment~$A$ guarantees that its resulting objective value $K(A,B^*)$ nearly matches the optimum value $K^*$. To verify this claim, the important observation is that for every client $j \in [n]$, our random sum of batch ends can be upper-bounded by
$$
\sum_{i\in[m]} {\cal E}^B_{i,A_i(j)} ~~=~~ \sum_{i \in [m]} {\cal E}^B_{i,R^{(j)}_{i}} ~~=~~ K(R^{(j)}) ~~\leq~~ (1 + 29\eps) \cdot K^*,
$$
where the last inequality holds since $R^{(j)}$ is a valid configuration. Therefore,
\[ K(A,B^*) ~~=~~ \max_{j \in [n]} \sum_{i\in[m]} {\cal E}^B_{i,A_i(j)} ~~\leq~~ (1 + 29\eps) \cdot K^* \ . \]

\paragraph{Translating $\boldsymbol{(A,B^*)}$ to a schedule.} That said, the assignment $A$ can still overflow the capacity of one or more batches. Moreover, in contrast to Section~\ref{sec:PTAS}, we do not know how to bound such overflows by an additive factor, but rather by a multiplicative factor. To differentiate between these notions, instead of considering $\Delta$-feasibility, we say that $A$ is a  $\sigma$-stretched assignment when $\sum_{j \in A_i^{-1}(b) } p_j  \leq \sigma \cdot B^*_{i,b}$ for every $i \in [m]$ and $b \in [\beta]$. 

To understand the implications of this new notion, the next claim describes the guarantees that can be obtained when a $\sigma$-stretched assignment is translated back to a standard schedule. Since the arguments involved are quite similar to those presented when proving Lemma~\ref{lem:AssignmentToSchedule}, the proof of this result is deferred to Appendix~\ref{app:ProofLemmaStrechedAssignment}.

\begin{lemma}
\label{lem:StrechedAssignmentToSchedule}%
Let $B$ be a batching, and let $A$ be a $\sigma$-stretched assignment with respect to~$B$. Then, $A$ can be translated in $O(nm)$ time to a schedule $\pi$ such that
$K(\pi) \leq \sigma \cdot K(A,B)$.
\end{lemma}

\subsection{Bounding the stretch of \texorpdfstring{$\boldsymbol{A}$}{}} \label{subsec:bound_stretch}

We conclude our analysis by showing that, with constant probability, the random assignment~$A$ is $(1+2\eps)$-stretched. As such, in conjunction with the translation method of Lemma~\ref{lem:StrechedAssignmentToSchedule}, we obtain a $(1+O(\eps))$-approximate schedule with constant probability, which can be arbitrary amplified via independent repetitions. 

\begin{lemma} 
\label{lem:RandAssignOverflow}
$A$ is a $(1+2\eps)$-stretched assignment with probability at least $1/2$.  
\end{lemma}

To establish this result, for every $i \in [m]$ and $b \in [\beta]$, let us define the random variable  $P_{i,b} = \sum_{ j \in A_i^{-1}(b) } p_j$, standing for the total processing time of the jobs assigned to batch $(i,b)$. With this definition, Lemma~\ref{lem:SingleBatchOverflow} below provides an upper bound of $\frac{1}{2m\beta}$ on the probability that $P_{i,b}$ exceeds the batch capacity $B^*_{i,b}$ by a factor of at least $1+2\eps$. Consequently, Lemma~\ref{lem:RandAssignOverflow} directly follows by applying the union bound.

\begin{lemma} 
\label{lem:SingleBatchOverflow}
$\prpar{ P_{i,b} \geq (1 + 2\eps) \cdot B^*_{i,b} } \leq \frac{1}{2m\beta}$, for every $i \in [m]$ and $b \in [\beta]$.
\end{lemma}
\begin{proof}
For every client $j \in [n]$ and configuration $c \in {\cal C}$, let $Z_{j,c} = \mathbbm{1} \{R^{(j)}=c \}$ be an indicator random variable for the event where configuration~$c$ was chosen for client~$j$. As a result, the random contribution of this client toward the total processing time $P_{i,b}$ of batch $(i,b)$ is given by  $Z_{i,j,b} = \sum_{c \in {\cal C}:c_i = b} p_j Z_{j,c}$. Noting that $P_{i,b} = \sum_{j \in [n]} Z_{i,j,b}$, this random variable can be decomposed into the contributions of large and small clients as follows:
\begin{equation} \label{eqn:decomp_Pib}
P_{i,b} ~~=~~ \sum_{j \in S} Z_{i,j,b} + \sum_{j \in L} Z_{i,j,b}  ~~=~~  \sum_{j \in S} Z_{i,j,b} + \sum_{j \in L} \sum_{\MyAbove{ c \in {\cal C}: }{ c_i = b }} p_j  x^*_{j,c}.    
\end{equation}
Here, the second equality holds since $x^*$ satisfies constraint~\eqref{three}, due to its feasibility in \mylp. Hence,  for every large client $j \in L$, we have $x^*_{j,c^{*(j)}}=1$ and $x^*_{j,c}=0$ for any $c \neq c^{*(j)}$, meaning that $Z_{j,c} = x^*_{j,c}$ with probability~1 for all $c \in {\cal C}$.

Given these observations, we  derive an upper bound on $\prpar{ P_{i,b} \geq (1 + 2\eps) \cdot B^*_{i,b} }$ by noting that 
\begin{eqnarray}
&& \pr{P_{i,b} \geq (1 + 2\eps) \cdot B^*_{i,b} } \nonumber \\
&& \qquad =~~ \pr{ \sum_{j \in S} Z_{i,j,b} \geq (1 + 2\eps) \cdot B^*_{i,b} - \sum_{j \in L} \sum_{\MyAbove{ c \in {\cal C} : }{ c_i = b }} p_j x^*_{j,c} } \label{eqn:SingleBatchOverflow_1}  \\
& & \qquad \leq~~ \pr{ \sum_{j \in S} Z_{i,j,b} \geq (1 + \eps) \cdot \sum_{j \in S} \sum_{\MyAbove{ c \in {\cal C}: }{ c_i = b }} p_j x^*_{j,c} + \eps B^*_{i,b} } \label{eqn:SingleBatchOverflow_2} \\
& & \qquad \leq~~ \pr{ \sum_{j \in S} \frac{ Z_{i,j,b} }{ \Lambda } \geq (1 + \eps) \cdot \left( \sum_{j \in {S}} \ex{ \frac{ Z_{i,j,b} }{ \Lambda } } + \frac{ \eps B^*_{i,b} }{ 2\Lambda } \right) }. \label{eqn:SingleBatchOverflow_3}
\end{eqnarray}
Here, equation~\eqref{eqn:SingleBatchOverflow_1} is obtained by plugging in decomposition~\eqref{eqn:decomp_Pib}. Inequality~\eqref{eqn:SingleBatchOverflow_2} holds since $\sum_{j \in [n]} \sum_{c\in {\cal C}: c_i = b } p_{j} x_{j,c}^* \leq B^*_{i,b}$, by constraint~\eqref{second}. Finally, inequality~\eqref{eqn:SingleBatchOverflow_3} follows by noting that $\expar{Z_{i,j,b} } = \sum_{c\in {\cal C}: c_i = b } p_j x^*_{j,c}$. To bound the resulting term, we employ the Chernoff-Hoeffding bound of Lemma~\ref{lem:Chernoff-Hoeffding} with respect to $\{ \frac{ Z_{i,j,b} }{ \Lambda }\}_{j \in S}$, which are indeed independent and $[0,1]$-bounded, since $Z_{i,j,b} \leq \Lambda$ for every small client~$j$. As a result,
\begin{eqnarray*}
& & \pr{ \sum_{j \in S} \frac{ Z_{i,j,b} }{ \Lambda } \geq (1 + \eps) \cdot \left( \sum_{j \in {S}} \ex{ \frac{ Z_{i,j,b} }{ \Lambda } } + \frac{ \eps B^*_{i,b} }{ 2\Lambda } \right) } \\
& & \qquad \leq~~ \exp \left( - \frac{ \eps^2 }{ 3 } \cdot \left( \sum_{j \in {\cal S}} \ex{ \frac{ Z_{i,j,b} }{ \Lambda }} + \frac{ \eps B^*_{i,b} }{ 2\Lambda } \right) \right) \\
& & \qquad \leq~~ \exp \left( - \frac{ \eps^6P }{ 6\Lambda } \right) \\
& & \qquad =~~ \frac{ 1 }{ 2m\beta} \ . 
\end{eqnarray*}
Here, the second inequality follows by noting that $\{Z_{i,j,b} \}_{j \in S}$ are non-negative random variables, and that $B^*_{i,b} \geq \min {\cal B} = \eps^3 P$. The final equality is obtained by plugging in $\Lambda = \frac{ \eps^6 P }{ 6 \ln(2m\beta)}$. 
\end{proof}
\section{Concluding Remarks}
\label{sec:summary}%

We conclude this paper by highlighting a number of compelling avenues for future research. Broadly speaking, we believe that our results lay the groundwork for investigating alternative objective functions as well as various machine settings integrated within  the repetitive scheduling framework. In relation to the specific outcomes of Theorems~\ref{thm:2Approx}-\ref{thm:QPTAS}, we proceed by posing three challenging questions, intended to complement some of our findings or to enhance their performance guarantees.

\paragraph{Avoiding an exponentially-sized LP.} As explained in Section~\ref{sec:2_approx_general}, our $2$-approximation for the day-dependent model relies on solving an exponentially-sized LP. While the latter admits a polynomial-time separation oracle, it would be interesting to examine whether an analogous result can be attained via a polynomially-sized LP, or even through a purely combinatorial algorithm. Toward this objective, one candidate approach could reside within the work of Plotkin et al.~\cite{shab}, who presented a polynomially-sized ILP. Can a fractional solution in this context be efficiently rounded to a constant-factor approximate schedule?

\paragraph{EPTAS for $\boldsymbol{O(1)}$ days.} As stated in Theorem~\ref{thm:PTAS}, it turns out that the day-dependent model admits an $O(n^{\tilde{O}(m / \eps^2)})$-time approximation scheme. Given some of the technical developments involved, it is natural to ask whether this finding could be refined, by proposing an EPTAS for constantly-many days. The latter corresponds to an approximation scheme that can be implemented in  $f(\frac{1}{\eps},m) \cdot n^{O(1)}$ time, for some function $f$. The main reason why our approach does not produce an EPTAS is its dynamic programming procedure, where rounded-down processing times have led us to consider $n^{\tilde{O}(m / \eps^2)}$ system states. Can one come up with a more compact formulation?

\paragraph{Hardness results for day-invariant instances.} As mentioned in Section~\ref{subsec:PreviousWork}, the day-dependent model is NP-hard to approximate within factor strictly better than $38/37$~\cite{hermelin2025fairness}. By contrast, despite our best efforts, we still do not know whether the day-invariant model is NP-hard or not, which we therefore pose as a particularly perplexing question for future research. Interestingly, when the number of days $m$ is given in its binary specification, it is unclear whether this problem even resides in $\mathrm{NP}$, since one cannot exclude the possibility that any optimal schedule may be comprised of $\Omega(m)$-many different daily schedules.

\bibliographystyle{abbrvnat}
\bibliography{BIB-Repetitive}

\newcommand{\noopsort}[1]{}
\begin{thebibliography}{43}
\providecommand{\natexlab}[1]{#1}
\providecommand{\url}[1]{\texttt{#1}}
\expandafter\ifx\csname urlstyle\endcsname\relax
  \providecommand{\doi}[1]{doi: #1}\else
  \providecommand{\doi}{doi: \begingroup \urlstyle{rm}\Url}\fi

\bibitem[Afrati et~al.(1999)Afrati, Bampis, Chekuri, Karger, Kenyon, Khanna,
  Milis, Queyranne, Skutella, Stein, and Sviridenko]{Foto1999}
F.~N. Afrati, E.~Bampis, C.~Chekuri, D.~R. Karger, C.~Kenyon, S.~Khanna,
  I.~Milis, M.~Queyranne, M.~Skutella, C.~Stein, and M.~Sviridenko.
\newblock Approximation schemes for minimizing average weighted completion time
  with release dates.
\newblock In \emph{Proceedings of the 40th Annual IEEE Symposium on Foundations
  of Computer Science}, pages 32--43, 1999.

\bibitem[Agnetis et~al.(2025)Agnetis, Billaut, Pinedo, and
  Shabtay]{AgnetisBPS25}
A.~Agnetis, J.~Billaut, M.~L. Pinedo, and D.~Shabtay.
\newblock Fifty years of research in scheduling -- {T}heory and applications.
\newblock \emph{European Journal of Operational Research}, 327\penalty0
  (2):\penalty0 367--393, 2025.

\bibitem[Alon and Spencer(2016)]{AlonS2016}
N.~Alon and J.~H. Spencer.
\newblock \emph{The Probabilistic Method}.
\newblock John Wiley \& Sons, fourth edition, 2016.

\bibitem[Arunachaleswaran et~al.(2022)Arunachaleswaran, Barman, and
  Rathi]{Arunachaleswaran2022}
E.~R. Arunachaleswaran, S.~Barman, and N.~Rathi.
\newblock Fully polynomial-time approximation schemes for fair rent division.
\newblock \emph{Mathematics of Operations Research}, 47\penalty0 (3):\penalty0
  1970--1998, 2022.

\bibitem[Babaioff and Feige(2025)]{Babaioff2025}
M.~Babaioff and U.~Feige.
\newblock Share-based fairness for arbitrary entitlements.
\newblock In \emph{Proceedings of the 57th Annual ACM Symposium on Theory of
  Computing}, pages 1544--1555, 2025.

\bibitem[Bansal et~al.(2006)Bansal, Caprara, and Sviridenko]{Bansal2006}
N.~Bansal, A.~Caprara, and M.~Sviridenko.
\newblock Improved approximation algorithms for multidimensional bin packing
  problems.
\newblock In \emph{Proceedings of the 47th Annual IEEE Symposium on Foundations
  of Computer Science}, pages 697--708, 2006.

\bibitem[Bertsimas et~al.(2011)Bertsimas, Farias, and
  Trichakis]{bertsimas2011fairness}
D.~Bertsimas, V.~F. Farias, and N.~Trichakis.
\newblock The price of fairness.
\newblock \emph{Operations Research}, 59\penalty0 (1):\penalty0 17--31, 2011.

\bibitem[Bertsimas et~al.(2012)Bertsimas, Farias, and
  Trichakis]{bertsimas2012fair}
D.~Bertsimas, V.~F. Farias, and N.~Trichakis.
\newblock On the efficiency-fairness trade-off.
\newblock \emph{Management Science}, 58\penalty0 (12):\penalty0 2234--2250,
  2012.

\bibitem[Bertsimas et~al.(2022)Bertsimas, Gupta, and
  Paschalidis]{bertsimas2022fairness}
D.~Bertsimas, V.~Gupta, and I.~C. Paschalidis.
\newblock Fairness in optimization and learning.
\newblock \emph{INFORMS Journal on Optimization}, 4\penalty0 (3):\penalty0
  209--239, 2022.

\bibitem[Breugem and Van~Wassenhove(2022)]{Breugem2020}
T.~Breugem and L.~N. Van~Wassenhove.
\newblock The price of imposing vertical equity through asymmetric outcome
  constraints.
\newblock \emph{Management Science}, 68\penalty0 (1):\penalty0 7977--7993,
  2022.

\bibitem[Chekuri and Khanna(2001)]{Chekuri2001}
C.~Chekuri and S.~Khanna.
\newblock A {PTAS} for minimizing weighted completion time on uniformly related
  machines.
\newblock In \emph{Proceedings of the 28th International Colloquium on
  Automata, Languages, and Programming}, pages 858--861, 2001.

\bibitem[Chekuri and Khanna(2004)]{ChekuriK04}
C.~Chekuri and S.~Khanna.
\newblock Approximation algorithms for minimizing average weighted completion
  time.
\newblock In J.~Y.-T. Leung, editor, \emph{Handbook of Scheduling: Algorithms,
  Models, and Performance Analysis}, pages 11--1--11--30. Chapman and Hall/CRC,
  2004.

\bibitem[Chekuri et~al.(2001)Chekuri, Motwani, Balas, and Stein]{Chekuri}
C.~Chekuri, R.~Motwani, N.~Balas, and C.~Stein.
\newblock Approximation techniques for average completion time scheduling.
\newblock \emph{SIAM Journal on Computing}, 31:\penalty0 146--166, 2001.

\bibitem[Dubhashi and Panconesi(2009)]{DubhashiP09}
D.~P. Dubhashi and A.~Panconesi.
\newblock \emph{Concentration of Measure for the Analysis of Randomized
  Algorithms}.
\newblock Cambridge University Press, 2009.

\bibitem[Fairbrother et~al.(2020)Fairbrother, Zografos, and
  Glazebrook]{Fairbrother2020}
J.~Fairbrother, K.~G. Zografos, and K.~D. Glazebrook.
\newblock A slot-scheduling mechanism at congested airports which incorporates
  efficiency, fairness, and airline preferences.
\newblock \emph{Transportation Science}, 54\penalty0 (1):\penalty0 115--138,
  2020.

\bibitem[Goemans et~al.(2002)Goemans, Queyranne, Schulz, and Skutella]{Goemans}
M.~X. Goemans, M.~Queyranne, A.~S. Schulz, and M.~Skutella.
\newblock Single machine scheduling with release dates.
\newblock \emph{SIAM Journal on Discrete Mathematics}, 15:\penalty0 165--192,
  2002.

\bibitem[Gr\"{o}tschel et~al.(1981)Gr\"{o}tschel, Lov\'{a}tsz, and
  Schrijver]{GrotschelEtAl81}
M.~Gr\"{o}tschel, L.~Lov\'{a}tsz, and A.~Schrijver.
\newblock The ellipsoid method and its consequences in combinatorial
  optimization.
\newblock \emph{Combinatorica}, 1:\penalty0 169--197, 1981.

\bibitem[Hall et~al.(1997)Hall, Schulz, Shmoys, and Wein]{HallEtAl1997}
L.~A. Hall, A.~S. Schulz, D.~B. Shmoys, and J.~Wein.
\newblock Scheduling to minimize average completion time: Off-line and on-line
  approximation algorithms.
\newblock \emph{Mathematics of Operations Research}, 22\penalty0 (3):\penalty0
  513--514, 1997.

\bibitem[Hebbar et~al.(2024)Hebbar, Shen, and Yao]{HebbarSY2024}
A.~Hebbar, Y.~Shen, and K.~Yao.
\newblock Fair price discrimination.
\newblock In \emph{Proceedings of the 35th Annual ACM-SIAM Symposium on
  Discrete Algorithms}, pages 1234--1250, 2024.

\bibitem[Heeger et~al.(2023)Heeger, Hermelin, Mertzios, Molter, Niedermeier,
  and Shabtay]{heeger2021equitable}
K.~Heeger, D.~Hermelin, G.~B. Mertzios, H.~Molter, R.~Niedermeier, and
  D.~Shabtay.
\newblock Equitable scheduling on a single machine.
\newblock \emph{Journal of Scheduling}, 26\penalty0 (2):\penalty0 209--225,
  2023.

\bibitem[Heeger et~al.(2025)Heeger, Hermelin, Itzhaki, Molter, and
  Shabtay]{heeger2025interval}
K.~Heeger, D.~Hermelin, Y.~Itzhaki, H.~Molter, and D.~Shabtay.
\newblock Fair repetitive interval scheduling.
\newblock \emph{Algorithmica}, 87:\penalty0 1340--1368, 2025.

\bibitem[Hermelin et~al.(2025)Hermelin, Molter, Niedermeier, Pinedo, and
  Shabtay]{hermelin2025fairness}
D.~Hermelin, H.~Molter, R.~Niedermeier, M.~Pinedo, and D.~Shabtay.
\newblock Fairness in repetitive scheduling.
\newblock \emph{European Journal of Operational Research}, 323\penalty0
  (3):\penalty0 724--738, 2025.

\bibitem[Ibarra and Kim(2004)]{chekuri2004}
O.~H. Ibarra and C.~E. Kim.
\newblock On multidimensional packing problems.
\newblock \emph{SIAM Journal on Computing}, 33\penalty0 (4):\penalty0 837--851,
  2004.

\bibitem[Iwata et~al.(2001)Iwata, Fleischer, and Fujishige]{IwataFF01}
S.~Iwata, L.~Fleischer, and S.~Fujishige.
\newblock A combinatorial strongly polynomial algorithm for minimizing
  submodular functions.
\newblock \emph{Journal of the ACM}, 48\penalty0 (4):\penalty0 761--777, 2001.

\bibitem[Karmarkar and Karp(1982)]{Karmarkar82}
N.~Karmarkar and R.~M. Karp.
\newblock An efficient approximation scheme for the one-dimensional bin-packing
  problem.
\newblock In \emph{Proceedings of the 23rd Annual IEEE Symposium on Foundations
  of Computer Science}, pages 312--320, 1982.

\bibitem[Knop and Kouteck{\'y}(2022)]{KnopK22}
D.~Knop and M.~Kouteck{\'y}.
\newblock Scheduling kernels via configuration {LP}.
\newblock In \emph{Proceeding of the 30th Annual European Symposium on
  Algorithms}, pages 73:1--73:15, 2022.

\bibitem[Kulik et~al.(2023)Kulik, Mnich, and Shachnai]{Kulik2023}
A.~Kulik, M.~Mnich, and H.~Shachnai.
\newblock Approximations for vector bin packing via iterative randomized
  rounding.
\newblock In \emph{Proceedings of the 23rd Annual IEEE Symposium on Foundations
  of Computer Science}, pages 1366--1376, 2023.

\bibitem[Lawler et~al.(1977)Lawler, {Rinnooy Kan}, and Brucker]{LRKB1977}
E.~L. Lawler, A.~H.~G. {Rinnooy Kan}, and P.~Brucker.
\newblock Complexity of machine scheduling problems.
\newblock \emph{Annals of Discrete Mathematics}, 1:\penalty0 343--362, 1977.

\bibitem[Lenstra et~al.(1990)Lenstra, Shmoys, and Tardos]{Lenstra90}
J.~K. Lenstra, D.~B. Shmoys, and E.~Tardos.
\newblock Approximation algorithms for scheduling unrelated parallel machines.
\newblock \emph{Mathematical Programming}, 46:\penalty0 259--271, 1990.

\bibitem[Lipton et~al.(2004)Lipton, Markakis, Mossel, and Saberi]{Lipton2004}
R.~J. Lipton, E.~Markakis, E.~Mossel, and A.~Saberi.
\newblock On approximately fair allocations of indivisible goods.
\newblock \emph{Journal of the ACM}, 51\penalty0 (4):\penalty0 385--406, 2004.

\bibitem[Neria and Tzur(2024)]{Tzur2024}
G.~Neria and M.~Tzur.
\newblock The dynamic pickup and allocation with fairness problem.
\newblock \emph{Transportation Science}, 58\penalty0 (4):\penalty0 821--840,
  2024.

\bibitem[Orlin(2009)]{Orlin09}
J.~B. Orlin.
\newblock A faster strongly polynomial time algorithm for submodular function
  minimization.
\newblock \emph{Mathematical Programming}, 118\penalty0 (2):\penalty0 237--251,
  2009.

\bibitem[Phillips et~al.(1998)Phillips, Stein, and Wein]{Phillips}
C.~Phillips, C.~Stein, and J.~Wein.
\newblock Minimizing average completion time in the presence of release dates.
\newblock \emph{Mathematical Programming}, 82:\penalty0 199--223, 1998.

\bibitem[Pinedo(2022)]{Pinedo12scheduling}
M.~L. Pinedo.
\newblock \emph{Scheduling: Theory, Algorithms, and Systems}.
\newblock Springer, sixth edition, 2022.

\bibitem[Plotkin et~al.(2026)Plotkin, Shabtay, and Fink]{shab}
A.~Plotkin, D.~Shabtay, and Y.~Fink.
\newblock Algorithms for fair repetitive scheduling based on total completion
  time criterion.
\newblock \emph{Computers \& Industrial Engineering}, 212:\penalty0 111659,
  2026.

\bibitem[Qi(2017)]{Qi17}
J.~Qi.
\newblock Mitigating delays and unfairness in appointment systems.
\newblock \emph{Management Science}, 63\penalty0 (2):\penalty0 566--583, 2017.

\bibitem[Queyranne(1993)]{Queyranne1993}
M.~Queyranne.
\newblock Structure of a simple scheduling polyhedron.
\newblock \emph{Mathematical Programming}, 58:\penalty0 263--268, 1993.

\bibitem[Schrijver(2000)]{Schrijver00}
A.~Schrijver.
\newblock A combinatorial algorithm minimizing submodular functions in strongly
  polynomial time.
\newblock \emph{Journal of Combinatorial Theory, Series B}, 80\penalty0
  (2):\penalty0 346--355, 2000.

\bibitem[Skutella(2001)]{DBLP:journals/jacm/Skutella01}
M.~Skutella.
\newblock Convex quadratic and semidefinite programming relaxations in
  scheduling.
\newblock \emph{Journal of the {ACM}}, 48\penalty0 (2):\penalty0 206--242,
  2001.

\bibitem[Smith(1956)]{Smith1956}
W.~E. Smith.
\newblock Various optimizers for single-stage production.
\newblock \emph{Naval Research Logistics Quarterly}, 3:\penalty0 59--66, 1956.

\bibitem[Svensson(2012)]{Svensson2012}
O.~Svensson.
\newblock Santa claus schedules jobs on unrelated machines.
\newblock \emph{SIAM Journal on Computing}, 41\penalty0 (5):\penalty0
  1318--1341, 2012.

\bibitem[Vardi and Haskell(2024)]{Vardi2025}
S.~Vardi and W.~Haskell.
\newblock The price of fairness of scheduling a scarce resource.
\newblock \emph{Operations Research}, 73\penalty0 (6):\penalty0 3104--3117,
  2024.

\bibitem[Verschae and Wiese(2014)]{VerschaeW14}
J.~Verschae and A.~Wiese.
\newblock On the configuration-{LP} for scheduling on unrelated machines.
\newblock \emph{Journal of Scheduling}, 17\penalty0 (4):\penalty0 371--383,
  2014.

\end{thebibliography}

\appendix
\section{Additional Proofs}

\subsection{Proof of Lemma~\ref{lem:RefinedStructureLemma}} 
\label{app:ProofRefinedStructure}

Let $\pi^*$ be an optimal solution to our day-invariant instance, which includes $m \leq \frac{\ln n} {\eps^3}$ days. The construction of the batching~$B^*$ and its corresponding assignment~$A^*$ is identical to the one described within the proof of Lemma~\ref{lem:StructureLemma}, other than having a different set of allowed batch capacities ${\cal B}$. Thus, we move directly to proving that $K(A^*,B^*) \leq (1+29\eps) \cdot K^*$, noting that this part is very similar as well.

To bound $K(A^*,B^*)$, we will argue below that for each client~$j \in [n]$ and day~$i \in [m]$, the end ${\cal E}^B_{i,A^*_i(j)}$ of the batch to which job~$(i,j)$ is assigned under~$A^*$ and~$B^*$ can be upper-bounded by
\begin{equation} 
\label{eqn:RefinedStructureLemmaProof}%
{\cal E}^B_{i,A^*_i(j)} ~~\leq~~ (1 + \eps)^2 \cdot C_{i,j}(\pi^*) + 13\eps P.
\end{equation}
Then, by summing this bound across all days, it follows that
$$
\sum_{i\in [m]} {\cal E}^B_{i,A^*_i(j)} ~~\le~~ (1+\eps)^2 \cdot \sum_{i \in [m]} C_{i,j}(\pi^*) + 13\eps mP ~~\leq~~ (1+3\eps) \cdot C_j(\pi^*) + 26\eps  K^*,
$$
where the last inequality holds since $K^* \geq \frac{mP}{2}$ by Lemma~\ref{lem:EnhancedLowerBound}. Therefore,
$$
K(A^*,B^*) ~~=~~ \max_{j \in [n]} \sum_{i\in [m]} {\cal E}^B_{i,A^*_i(j)} ~~\leq~~  (1+3\eps) \cdot \max_{j \in [n]}  C_j(\pi^*) + 26\eps  K^* ~~=~~ (1+29\eps)\cdot K^*.
$$
We prove inequality~\eqref{eqn:RefinedStructureLemmaProof} by considering two cases, depending on whether the job $(i,j)$ is assigned to an odd-indexed batch or to an even-indexed batch. 

\paragraph{Case 1: {$\boldsymbol{A^*_i(j)= b = 2\hat{\chi}-1}$} for some {$\boldsymbol{\hat{\chi} \leq \chi}$.}} In this case, we have
\begin{eqnarray}
{\cal E}^B_{i,A^*_i(j)} & = & \sum_{\hat{b} \leq b} B^*_{i,\hat{b}} \nonumber \\
&\leq& (1 + \eps) \cdot \sum_{\hat{b} \leq b} \sum_{k \in (A^*_i)^{-1}(\hat{b})} p_k + \eps^3 \beta P \label{oneone}\\
\label{twotow}
& \leq & (1 + \eps) \cdot \max_{ \MyAbove{k\in [n]:}{A^*_i(k) \leq b}} C_{i,k}(\pi^*) + 12\eps P \\
\label{threethree}
& \leq & (1 + \eps)^{\hat{\chi}} \cdot \eps^3  P + 12\eps P \\
\label{fourfour}
& \leq & (1 + \eps)^2 \cdot C_{i,j}(\pi^*) + 12\eps P.
\end{eqnarray}
Here, inequality~\eqref{oneone} holds since $B^*_{i,\hat{b}} = \lceil \sum_{k \in (A^*_i)^{-1}(\hat{b})} p_k \rceil^{(\mathcal{B})}$ for every $\hat{b} \in [\beta]$, and since $\lceil x \rceil^{(\mathcal{B})} \leq (1 + \eps) \cdot x + \eps^3 P$, by definition of $\lceil \cdot \rceil^{(\mathcal{B})}$ and $\mathcal{B}$. Inequality~(\ref{twotow}) follows by noting that $\sum_{\hat{b} \leq b} \sum_{k \in (A^*_i)^{-1}(\hat{b})} p_k$ coincides with the maximal completion time of any job assigned to batches $(i,1),\ldots,(i,b)$. In addition, we know that $\beta \leq 2|\mathcal{B}| \leq \frac{ 12 }{ \eps } \ln \frac{ 1 }{ \eps }$ by \eqref{eqn:RefinedGoodBatchingLength}. Inequality~(\ref{threethree}) is obtained by noting that, since $A^*_i(j) = 2\hat{\chi}-1$, the previously-mentioned maximal completion time is necessarily attained for one of the jobs in $(A^*_i)^{-1}(2\hat{\chi}-1)$, implying that it is upper-bounded by $(1 + \eps)^{\hat{\chi}-1} \cdot \eps^3 P$. Finally, inequality~(\ref{fourfour}) holds since $C_{i,j}(\pi^*) \geq (1 + \eps)^{\hat{\chi}-2} \cdot \eps^3 P$.

\paragraph{Case 2: $\boldsymbol{A^*_i(j)= b = 2\hat{\chi}}$ for some {$\boldsymbol{\hat{\chi} \leq \chi}$.}}
In this case, we know that $(A^*_i)^{-1}(b) = \{j\}$, meaning that $(i,j)$ was the only job assigned to batch $(i,b)$. Thus, we have $B_{i,b}^*=  \lceil p_j \rceil^{({\cal B})}
\leq (1+\eps) \cdot p_j + \eps^3 P$, and similarly to the analysis above,
\begin{eqnarray*}
{\cal E}^B_{i,A^*_i(j)} &=& \sum_{\hat{b} < b} B^*_{i,\hat{b}}+B^*_{i,b} \\
&\leq &  (1 + \eps) \cdot \max_{ \MyAbove{ k \in [n]:}{A^*_i(k) < b}} C_{i,k}(\pi^*) + 12\eps P +
(1+\eps) \cdot p_j + \eps^3  P \\
&\leq & (1+\eps) \cdot \left(\max_{ \MyAbove{ k \in [n]:}{A^*_i(k) < b}} C_{i,k}(\pi^*) + p_j \right) + 13\eps   P\\
&= & (1+\eps) \cdot C_{i,j}(\pi^*) + 13\eps P.
\end{eqnarray*}
To better understand the last equality, recall that the set of jobs that precede $(i,j)$ in $\pi^*_i$ is exactly those that were assigned to batches $(i,1),\ldots,(i,b-1)$, implying that $C_{i,j}(\pi^*)$ is precisely the sum of processing times of these jobs plus $p_j$. 


\subsection{Proof of Lemma~\ref{lem:StrechedAssignmentToSchedule}} \label{app:ProofLemmaStrechedAssignment}%

Our method for creating the schedule $\pi_i$ of each day $i \in [m]$ is identical to the one proposed within the proof of Lemma~\ref{lem:AssignmentToSchedule}. Specifically, given the batching $B$ and its $\sigma$-stretched assignment $A$, we begin by processing the set of jobs $A_i^{-1}( 1 )$ assigned to the batch $(i,1)$, in arbitrary order. Next, we process the set of jobs $A_i^{-1}( 2 )$ assigned to $(i,2)$, again in arbitrary order. The remaining sequence of sets $A_i^{-1}( 3 ), \ldots, A_i^{-1}( \beta )$ is handled along the same lines.

Letting $\pi=(\pi_1,\ldots,\pi_m)$ be the resulting schedule, we claim that the completion time $C_{i,j}(\pi)$ of each job $(i,j)$ is at most $\sigma \cdot {\cal E}^B_{i,A_i(j)}$. Indeed, given how $\pi_i$ is defined above, this completion time is upper-bounded by the total processing time of the jobs assigned to batches $(i,1),\ldots,(i,A_i(j))$, and therefore,
$$
C_{i,j}(\pi_i) ~~\leq~~ \sum_{\MyAbove{k\in[n]:}{A_i(k) \leq A_i(j)}} p_{i,k} ~~=~~ \sum_{b \leq A_i(j)} \sum_{ \MyAbove{k\in[n]:}{A_i(k) = b}} p_{i,k} ~~\leq~~ \sigma \cdot \sum_{b \leq A_i(j)} B_{i,b} ~~=~~ \sigma \cdot {\cal E}^B_{i,A_i(j)},
$$
where the second inequality holds since $A$ is a $\sigma$-stretched assignment. By summing the above inequality across all days, we conclude that $C_j(\pi) \leq \sigma \cdot \sum_{i \in [m]} {\cal E}^B_{i,A_i(j)} \leq \sigma \cdot K(A,B)$ for every client $j \in [n]$, meaning that $K(\pi) \leq \sigma \cdot K(A,B)$, as desired.

\end{document}